\documentclass{ws-ijmpc}

\usepackage{graphics}
\usepackage{url}
\usepackage{graphicx}
\usepackage{color}

\definecolor{orange}{rgb}{0.8,0.2,0.2}

\newcommand{\linguistic}{lexical innovation}
\newcommand{\linguistics}{lexical innovations}

\begin{document}

\markboth{Marco Alberto Javarone}
{Competitive dynamics of lexical innovations in multi-layer networks}

\catchline{}{}{}{}{}

%
\title{Competitive dynamics of lexical innovations in multi-layer networks}
\author{Marco Alberto Javarone}
\address{DUMAS - Dept. of Humanities and Social Sciences, 07100 Sassari, Italy \and Dept. of Mathematics and Computer Science, 09124 Cagliari, Italy}
\maketitle
%

\begin{abstract}
We study the introduction of \linguistics~into a community of language users. Lexical innovations, i.e., new terms added to people's vocabulary, play an important role in the process of language evolution.
Nowadays, information is spread  through a variety of networks, including, among others, online and offline social networks and the World Wide Web. The entire system, comprising networks of different nature, can be represented as a multi-layer network.
In this context, \linguistics~diffusion occurs in a peculiar fashion. In particular, a \linguistic~can undergo three different processes: its original meaning is accepted; its meaning can be changed or misunderstood (e.g., when not properly explained), hence more than one meaning can emerge in the population; lastly, in the case of a loan word, it can be translated into the population language (i.e., defining a new \linguistic~or using a synonym) or into a dialect spoken by part of the population. Therefore, \linguistics~cannot be considered simply as information.
We develop a model for analyzing this scenario using a multi-layer network comprising a social network and a media network. The latter represents the set of all information systems of a society, e.g., television, the World Wide Web and radio. Furthermore, we identify temporal directed edges between the nodes of these two networks. In particular, at each time step, nodes of the media network can be connected to randomly chosen nodes of the social network and vice versa. In so doing, information spreads through the whole system and people can share a \linguistic~with their neighbors or, in the event they work as reporters, by using media nodes.
Lastly, we use the concept of ``linguistic sign'' to model \linguistics, showing its fundamental role in the study of these dynamics.
Many numerical simulations have been performed to analyze the proposed model and its outcomes.
\end{abstract}
\ccode{PACS Nos.: 89.75.Hc, 89.65.-s, 89.75.Fb}

\section{Introduction}\label{intro}
The study of language has a strongly interdisciplinary nature, embracing among others linguistics, psychology, philosophy and computational sciences.  Computational tools in particular provide important information about the structure of  a  language~\cite{borge01}\cite{tenenbaum01}\cite{hill01}. 
Since a language can be construed as a complex system that evolves over time, many authors have shown that a statistical mechanics based framework can be used to analyze linguistics phenomena~\cite{loreto01}\cite{castello01}\cite{loreto03}\cite{castello02}.
The field of ``language dynamics''~\cite{loreto01}\cite{loreto02}, a relatively new research line,  deals with the emergence, evolution and extinction of languages.
Interestingly Wittgenstein can be considered a pioneer of language dynamics, as he developed the concept of language game~\cite{wittgenstein01}. 
Inspired by Wittgenstein's observations, Steels and collaborators developed one of the most famous language games, called the Naming Game~\cite{steels01}. It is based on local interactions among agents and allows one to simulate phenomena such  as the generation of a common vocabulary.
The Naming Game can be played on networks by representing interacting agents as connected nodes~\cite{baronchelli01}\cite{asta01}. Therefore, many aspects of this game are closely associated with the structure of the agents’ network. 
These kinds of language models focus on the concept of ``word'' as an atomic element. Though this choice simplifies some aspects of the underlying linguistic phenomena, it does need to be given  particular attention.
For instance, as shown in~\cite{javarone01}, the study of the emergence of acronyms relies on the concept of ``linguistic sign''~\cite{saussure01} (sign, hereinafter). Hence, in this case, the atomic elements are the signifier and the signified of a sign.
Nowadays, information is disseminated through different channels, e.g., people's interactions and technologies such as the television, the World Wide Web, and the radio, that make up the entire information system of a society. Complex networks and multi-layer networks in particular~\cite{dedomenico01} can be adopted to represent this scenario.
We identify two different kinds of communication: ``active'' and ``passive'', depending on the nature of the information channel. 
For example, a conversation between two people is active communication, whereas watching television is passive communication.
The main difference between the two is that in active communication people can interact with one another immediately, while this not so in passive communication. Hence, in the event that a conversation between people contains something that is not clear, the listener can immediately ask the speaker for further explanations. 
Active communication can take the form of face-to-face interactions, phone calls and online services for instant messaging. 
On the other hand, watching the television or reading newspapers, are examples of passive communication.
It is important to note that, if a piece of information is not clearly explained, then it may be misunderstood when spread by passive communication since, as discussed above, listeners (or readers) cannot ask for additional explanations immediately.
For instance, during the recent financial and economic crisis that hit Italy (and many other countries), the English term ``spread'' was frequently used by the national media. This term has many meanings in the English language but has now gained wide usage in Italy (without translating it into Italian) to indicate the difference in yield between Italian and German long term government bonds. The media rarely provided an accurate explanation of this concept, the result being that people became familiar with the term "spread'' and sometimes used it without knowing what it actually meant. 
In the light of these considerations, active communication allows agents to provide both messages (i.e., information) and explanations (i.e., meanings), while passive communication only provides messages, thus listeners/readers have to know the meaning of each word to fully understand them.
In this work we study the dynamics of \linguistics~diffusion in a community of language users. We represent the entire information system by a single network of media nodes (i.e., web-sites, television, radio, etc.). Since nodes of the social network interact with those of the media network, and vice-versa, these two networks can be represented as a multi-layer network.
We identify temporal edges between media nodes and nodes of the social networks (i.e., people), and vice-versa. In particular, media nodes have a readership/audience, whereas people acting as reporters can use the media network to spread their information.
In the proposed model, interactions among people are viewed as active communication, whereas interactions between people and media nodes as passive communication. Active communications are mapped to undirected edges while passive communications are mapped to directed edges (i.e., arrows).
We analyze the peculiar behavior of the diffusion of \linguistics, showing the emergence of competitive dynamics.
The remainder of the paper is organized as follows: Section~\ref{sec:language} gives a brief linguistic overview of \linguistics~and the linguistic sign. Section~\ref{sec:model} describes the proposed model for studying the diffusion of \linguistics. Section~\ref{sec:results} provides the results of numerical simulations. Section~\ref{sec:conclusions} ends the paper. 
\section{Language as a complex system: a brief Linguistic overview} \label{sec:language}
A language evolves over time as the result of a variety of phenomena, e.g., the evolutionary processes of phonetics, syntax and lexicon. Of these phenomena, \linguistics~have a significant role and often an aesthetic effect. In general, \linguistics~are new terms adopted by a language for different needs.
Every \linguistic~originates from a source, which introduces it into the language of a population, with a view to its diffusion. Making a guess about the adoption of a \linguistic~is not a trivial task.
The reason for this is the large amount of sometimes unknown variables involved.
To study these dynamics, the concept of sign~\cite{saussure01} has an important role. The sign, the basic unit of language, is composed of a signifier and a signified. The signifier is the shape of a word, whereas the signified represents its meaning. The signified  is related to a referent, i.e., the actual object. These three entities (i.e., signifier, signified and referent) are vertices of the ``Triangle of reference''~\cite{ogden01}, and the sign is identified on the side connecting signifier and signified.
From a computational perspective, the sign can be described by the following relation: $Sign = (Signifier, Signified)$.
Hence, when more signifieds are associated with the same signifier, every pair $(signifier, signified)$ constitutes a sign, that in turn can survive (i.e., be used by a population) or not over time.
In general, the vocabulary of a language includes homonyms and homographs, i.e., words that are pronounced or written in the same way but have a different meaning. These words are signs with the same signifier but different signified. For example, ``bear'' means ``to carry'' (verb) or refers to the animal (noun).
During the process of language evolution, a word may acquire a new meaning (i.e., signified). For instance, the Latin word ``captivus'' originally meant prisoner'',  but later its meaning changed to ``bad person'' (e.g., in Italian, a "modern" Latin language, the noun ``cattivo'' means ``bad person''). Other examples are the sign ``tweet'' and the sign ``glass''. Nowadays, the former has acquired two main meanings: the ``chirping sound of a bird'' and a ``short message sent by the social network Twitter''. Instead, the sign glass has many meanings, e.g., it refers to the material and also to the container for drinks.
These examples show that, as stated before, two or more signifieds for the same sign can coexist, neither one prevailing.
The concept of linguistic sign is very important to model the linguistic processes that a \linguistic~undergoes. In particular, when a \linguistic~is introduced into a community, three possible processes  can occur: \textit{L1}) the \linguistic~is accepted, therefore both the signifier and the signified are not altered by anyone; \textit{L2}) Other signifieds emerge as a result of misunderstandings or of modifications; \textit{L3}) Some people decide to use a synonym or, in the event of loan words, someone translates the \linguistic~into the native language (i.e., the population language) or into a dialect known to part of the population.
From a computational perspective, these three processes can be formalized as follows:
\begin{itemize}
\item \textit{L1}: $Sign = (\bar{x}, \bar{y}) \to Sign = (\bar{x}, \bar{y})$
\item \textit{L2}: $Sign = (\bar{x}, \bar{y}) \to Sign = (\bar{x}, Y)$
\item \textit{L3}: $Sign = (\bar{x}, \bar{y}) \to Sign = (X, \bar{y})$
\end{itemize}
\noindent each \linguistic~$Sign = (\bar{x}, \bar{y})$ changes depending on the process (i.e., \textit{L1},\textit{L2},\textit{L3}), the elements $\bar{x}$ and $\bar{y}$ represent the signifier and the signified, respectively, of a \linguistic. Lastly, $X$ represents the set of $N$ different signifieds and $Y$ the set of $N$ different signifiers.
For instance, if we consider the Italian language, the \linguistic~``tweet'' with the meaning related to the social network (i.e., Twitter) underwent the process \textit{L1}, i.e., Italian people use it with that meaning without translating it. On the other hand, the \linguistic~``spread'' has not yet been translated into Italian, but has different meanings as a result of misunderstandings (i.e., \textit{L2}), whereas the \linguistic~``body'' now commonly used in Italy refers to a ladies' body suit, but in this case is not the result of a misunderstanding (i.e., \textit{L2}). Instead, \linguistics~such as ``computer'' can be translated into ``calcolatore'' or ``elaboratore elettronico''. The third process (i.e., \textit{L3}) is very common  in the French language, due to cultural traditions, as French people do not like to use loan words (e.g., ``computer'' is often called ``ordinateur'' and the tool ``mouse'' is called ``souris''). 
Moreover, the third process also takes place where dialects exist, i.e., varieties of a language that are characteristic of a small group of a linguistic community.
\section{\textit{Lexical innovations} on multi-layer networks }\label{sec:model}
\begin{figure}[!ht]
\centering
\includegraphics[width=3.0in]{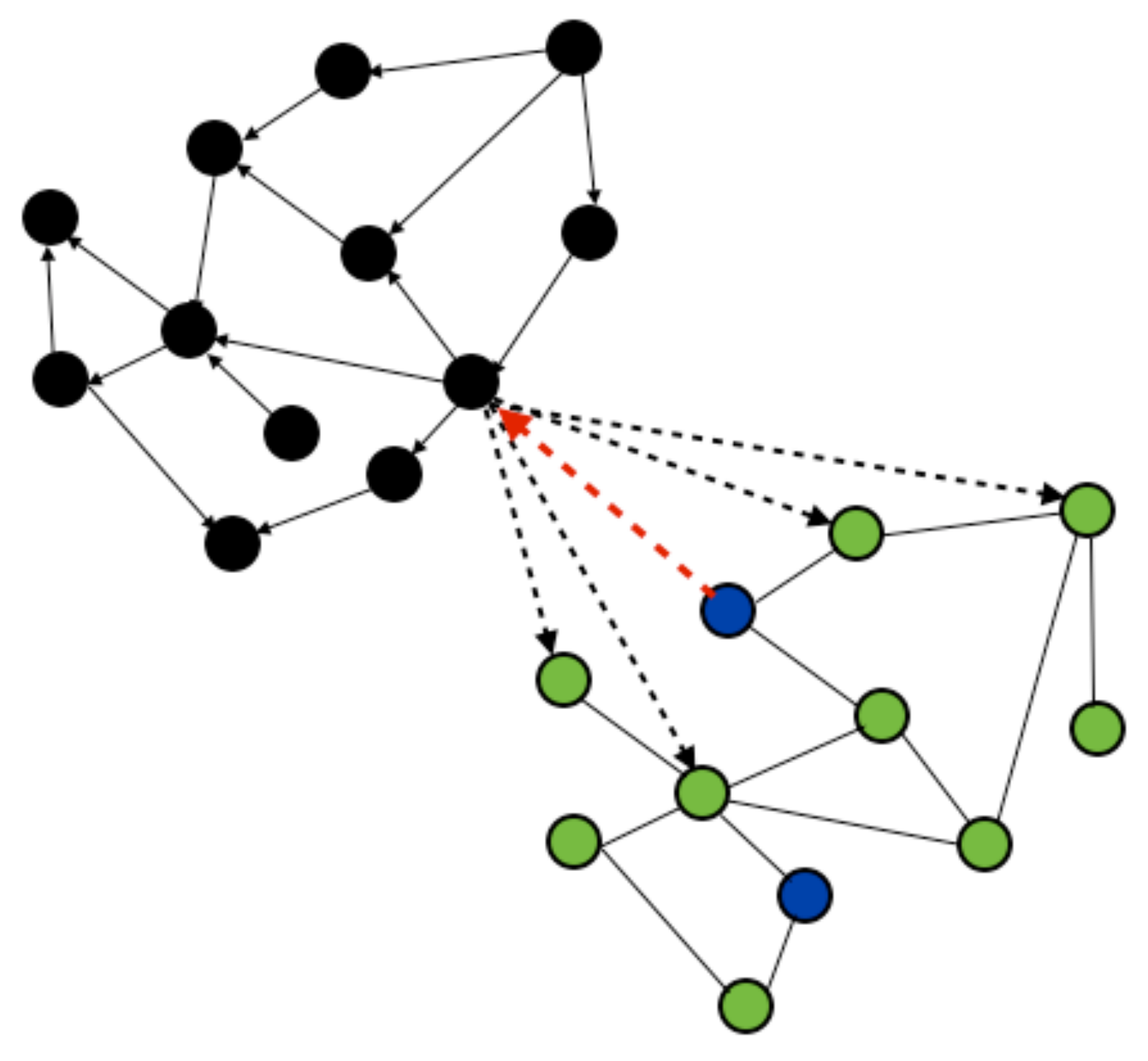}
\caption{\label{fig:model} Multi-layer network of the proposed model: an undirected social network (connections represented by solid black lines), and a directed media network (connections represented by solid black arrows). Black nodes denote media nodes, green and blue nodes agents of the social network. In particular, blue nodes represent reporters. Dashed black arrows are temporal interactions between media nodes and agents, whereas the dashed red arrow is a temporal connection between reporter and a media node. 
}
\end{figure}

Let us now introduce our model to study the diffusion dynamics of \linguistics~in a community of language users. We consider a multi-layer network composed of two layers: an undirected social network, where people (agents hereinafter) interact sharing information with their neighbors, and a directed media network, where media nodes share information via technologies such as web services~\cite{w301}. 
Furthermore, since agents can receive information from the media network, we identify temporal directed edges between media nodes and randomly chosen nodes of the social network. Each media node can generate a number of temporal directed edges~\cite{saramaki01} in accordance with its audience/readership.
On the other hand, agents working as reporters can share information via a media node. Hence, at each time step, a temporal directed edge is identified between reporters and randomly chosen media nodes --see Figure~\ref{fig:model}.

In the proposed model, a randomly selected agent invents a \linguistic~and initiates its diffusion. In so doing, at each time step, all agents who know the \linguistic, share it with one of their (randomly selected) neighbors or, in the event they work as reporter, use a media node to share it.
In turn, media nodes that receive the \linguistic~generate a number of connections (i.e., temporal directed edges) with (randomly chosen) agents for onward transmission.
Let us recall that \linguistics~are signs composed of a signifier and a signified. Moreover, we observe that in real interactions between people and media people only receive signifiers. Therefore, as mentioned above, when media nodes send a \linguistic~to their listeners/readers, it is possible that some misunderstandings occur. 
In particular, listeners /readers who do not know the \linguistic~try to identify the related signified by considering the context in which it is used. Hence, in the event the identified signified is different from the original one, a new sign is generated.
Furthermore, it may also happen that listeners/readers, who understand the signified of the \linguistic, decide to use it anyway by assigning a new signified. 
In so doing, the proposed model represents the first two above described linguistic processes (i.e., \textit{L1} and \textit{L2}), (see Section~\ref{sec:language}), undergone by a \linguistic~introducing the following simple rule: if an agent receives an unknown \linguistic~, he/she invents a new signified. Note that the term ``invent'', in this context, summarizes the process of identifying an appropriate signified by considering the context in which a \linguistic~is used. Hence, for each \linguistic~more than one sign can be generated. 
In general, the success of a sign depends on both the interactions among agents using it and on other factors such as its appeal (e.g., in terms of sound or style). 
Therefore, we introduce two parameters: number of confirmations $\sigma$ and fitness $\nu$. The number of confirmations considers successful interactions between agents, i.e., both agents use the same signified whereas fitness represents the probability of each sign surviving in the social network, i.e., its appeal or similar factors. 
 When two agents, let's say $x$ and $y$, who know the \linguistic~interact, they compare their signifieds. 
Now, in the event that both agents use the same signified, they increase their own number of confirmations. In the opposite case, they decide which signified to save by performing a weighted random selection between the success probabilities of each agent, $W_x$ and $W_y$, respectively. In particular, the success probability $W_x$ (or $W_y$) is computed as follows:
\begin{equation} \label{eq:signified_comparison}
W_x = \frac{\sigma_{x}\nu_{k}}{\sigma_{x}\nu_{k}+\sigma_{y}\nu_{z}} 
\end{equation}
\noindent with $\sigma_{x}$ and $\sigma_{y}$ number of confirmations received by agent $x$ and agent $y$, respectively. The parameters $\nu_{k}$ and $\nu_{z}$ are the fitness of the signifieds used by $x$ and by $y$, respectively. 
Note that the survival probability of an agent depends on a global parameter $\nu$ and on an individual parameter $\sigma$. The latter is computed by the agent him/herself considering personal experience, i.e., the number of times he/she found another agent with the same signified.
As a consequence, only signifieds with numerous confirmations and/or with a high fitness value survive over time.
To summarize, the proposed model comprises the following steps:
\begin{enumerate}[(1)]
\item Define two networks: the social network and the media network.
\item Add a \linguistic~(signifier, signified) to the vocabulary of a randomly selected agent. The original signified has (randomly computed) fitness $\nu_0$. The selected agent starts the spreading process sharing the \linguistic~with one (randomly chosen)neighbor.
\item Each agent, who knows the \linguistic, shares it as follows:
\begin{enumerate}[a)]
\item IF the $x$th agent is not a reporter, he/she shares the \linguistic~with one (randomly chosen) neighbor:
\begin{enumerate}[i.]
\item The selected neighbor (e.g., the $y$th agent) knows the \linguistic, hence agents play as follows:
\begin{itemize}
\item in the event both agents have the same signified, they increase their individual number of confirmations, i.e., $\sigma_x$ and $\sigma_y$.
\item in the event they have a different signified, a weighted random selection is performed between the success probabilities, $W_x$ and $W_y$, to decide which signified should be saved by both agents. Values of $W_x$ and $W_y$ are computed by Eq~(\ref{eq:signified_comparison}).
\end{itemize}
\item The selected neighbor does not know the \linguistic, hence adds it to its vocabulary.
\end{enumerate}
\item ELSE he/she uses a media node to share the \linguistic: the media node generates $n$ temporal directed edges with agents to send them the \linguistic. All agents who receive the \linguistic~from media nodes, and do not know it, invent a new signified (with a randomly computed fitness $\nu_x$).
\end{enumerate} 
\item Repeat from (3) until all agents know the \linguistic~and use a common signified.
\end{enumerate}
In principle, it may happen that more than one signified survives in the social network, i.e., more than one sign with the same signifier coexist over time. Therefore, we define a maximum number of iterations after which the algorithm terminates. 
Finally, let us recall that the same signifieds (codified as numbers), of a \linguistic, have the same fitness value.
\section{Simulations}\label{sec:results}
We performed numerical simulations of the proposed model using different structures of the agent network, e.g., scale-free and small-world, and varying the population size $N$. All agent networks have been generated with an average degree $\langle k \rangle = 6$.
In each population, $10\%$ of $N$ works as reporters, hence they can use the media network to spread the \linguistic. 
In turn, the media network has an Erd\"{o}s-Renyi graph (E-R graph hereinafter)~\cite{erdos01} structure with an average degree $\langle k \rangle = 2$. The structure of the media network is not important as, in the proposed model, the most important interactions are those among agents and between agents and media nodes.
The main role of the media network is to rapidly spread the \linguistic~in the agents’ network, as each media node has an audience (or readership). 
The audience is randomly assigned to each media node and can have a value in the range $[1, 25\%$ of $N$].
We limited the maximum number of signifieds, invented by agents, to four. This limitation is explained by the fact that, as people hear/read a new sign (i.e., a \linguistic~or a word they do not know), they usually try to identify its signified (i.e., its meaning) by considering the phrase or the context in which that sign is used. Therefore, the number of possible signifieds can be substantially reduced. 
Furthermore, since each signified of the \linguistic~has a fitness, we defined three sets of fitness values: $A = \{\nu_0 = 0.75, \nu_1 = 0.5, \nu_2 = 0.25, \nu_3 = 0.1\}$, $B = \{\nu_0 = 0.1, \nu_1 = 0.25, \nu_2 = 0.5, \nu_3 = 0.75\}$ and $C = \{\nu_0 = 0.65, \nu_1 = 0.7, \nu_2 = 0.7, \nu_3 = 0.65\}$.
Fitness sets $A$ and $B$ are equivalent because both contain four different values, in descending and ascending order, respectively. Instead, the set $C$ contains similar fitness values.
Results of simulations performed varying the fitness sets made it possible to evaluate how fitness affects the outcomes of the proposed model.
Lastly, for the sake of completeness, we also give a brief analysis of the model, performed without imposing a maximum number on the signifieds invented by agents. 

\subsection{Spreading dynamics}
We analyze the diffusion of \linguistics~over time on different agent networks.
This phenomenon is not trivial as it involves two main mechanisms: sharing a \linguistic~and convergence on common signifieds.
Since the proposed model considers ``active'' communication (i.e., both the signifier and the signified are communicated) among agents and ``passive'' communication (i.e., only the signifier is communicated) between agents and media nodes, we recall that more than one signified can be generated from the same \linguistic~.
We study this phenomenon by analyzing the density of agents who add a \linguistic~to their vocabulary, considering each invented signified. In so doing, it is possible to evaluate whether one or more signifieds prevail in the population.
\subsubsection*{Fully-connected networks}
\begin{figure}[!ht]
\centering
\includegraphics[width=3.6in]{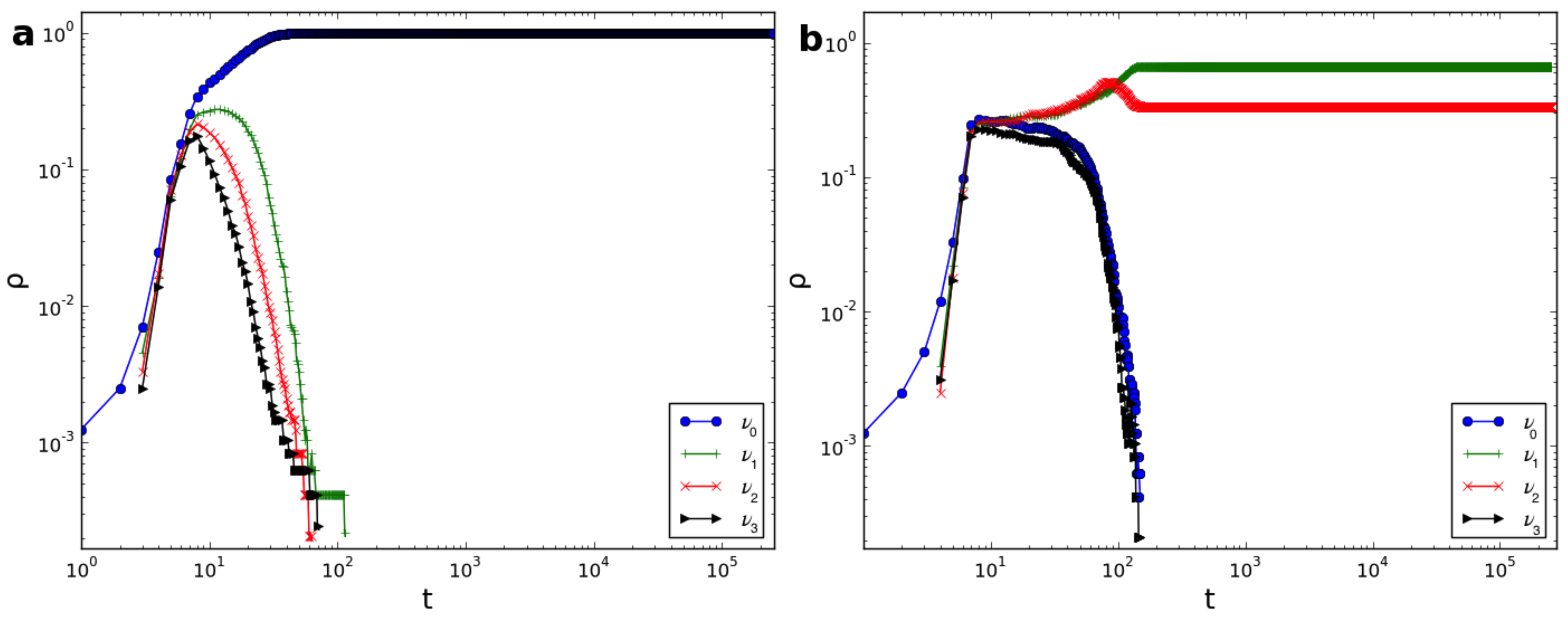}
\caption{\label{fig:density_fc} Density of agents who add the \linguistic, over time, in a FC network with $N=1600$ (results are averaged over $10$ simulations). As indicated in the legend, each curve refers to a different signified, having fitness $\nu$. \textbf{a} Results for fitness set $A$. \textbf{b} Results for fitness set $C$.}
\end{figure}

A fully-connected (FC hereinafter) network allows each agent to interact with the entire population. In principle, this structure can represent a real scenario where people communicate via face-to-face interactions and online social networks, and every person has the opportunity of getting in touch with all the others.
Figure~\ref{fig:density_fc} illustrates the density of agents $\rho$ who add the \linguistic~to their vocabulary over time.
\subsubsection*{Scale-Free Networks}
\begin{figure}[!ht]
\centering
\includegraphics[width=3.6in]{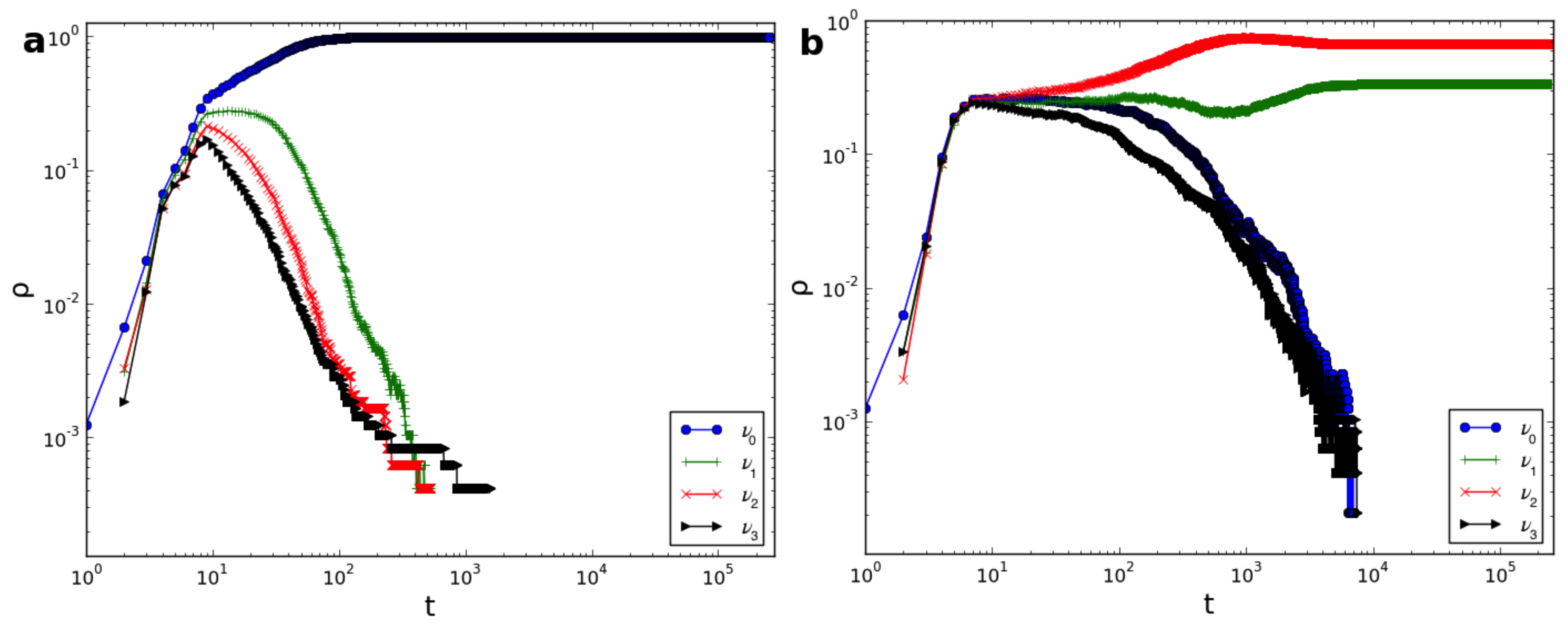}
\caption{\label{fig:density_sf} Density of agents who add the \linguistic, over time, in a SF network with $N=1600$ (results are averaged over $10$ simulations). As indicated in the legend, each curve refers to a different signified, having fitness $\nu$. \textbf{a} Results for fitness set $A$. \textbf{b} Results for fitness set $C$.}
\end{figure}

The scale-free (SF hereinafter) networks have been generated using the Barabasi-Albert model~\cite{barabasi01} with $m=3$, hence the scaling parameter of their degree distribution $P(k)$ is $\gamma \sim 3$. Note that SF networks are characterized by the presence of a few high degree nodes, called hubs and numerous low degree nodes.
Figure~\ref{fig:density_sf} illustrates the density of agents $\rho$ who add the \linguistic~to their vocabulary over time.
\subsubsection*{Small-World Networks}
\begin{figure}[!ht]
\centering
\includegraphics[width=3.6in]{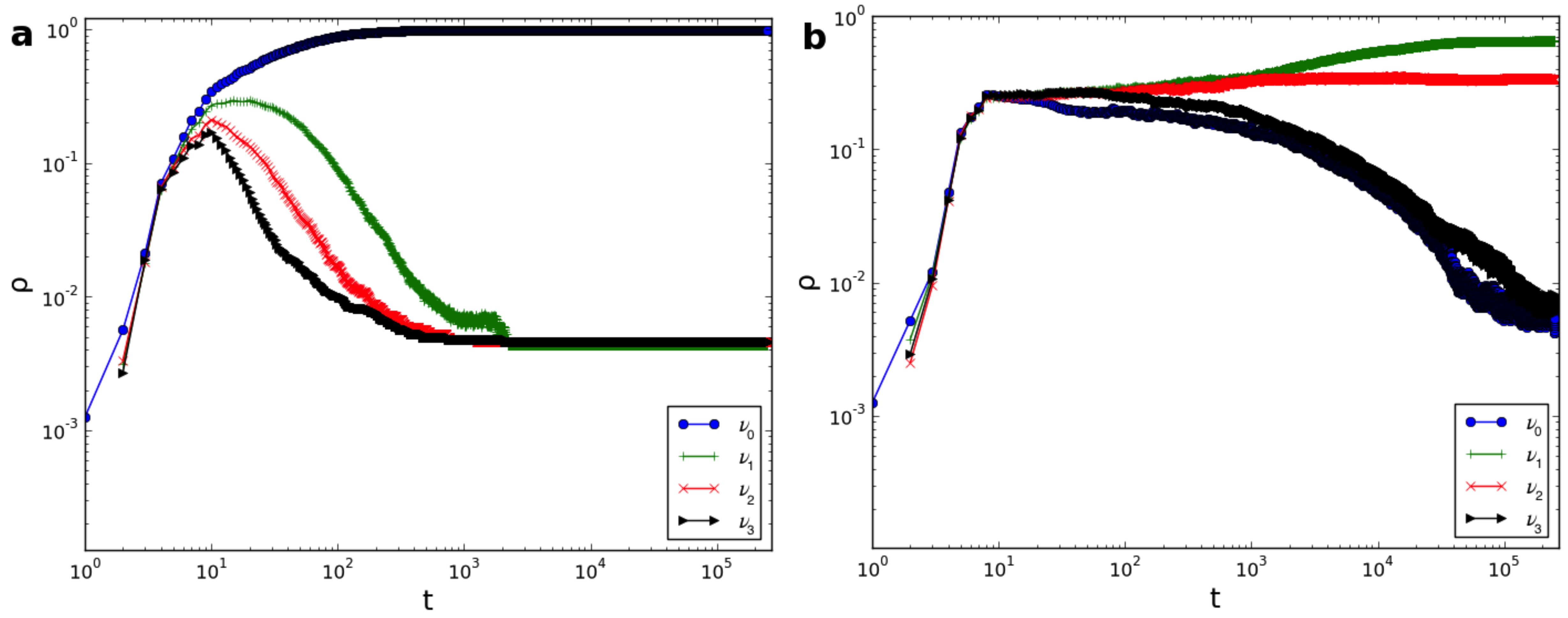}
\caption{\label{fig:density_ws_1} Density of agents who add the \linguistic, over time, in networks generated by the WS model with $\beta = 1.0$ and $N=1600$ (results are averaged over $10$ simulations). As indicated in the legend, each curve refers to a different signified, having fitness $\nu$. \textbf{a} Results for fitness set $A$. \textbf{b} Results for fitness set $C$.}
\end{figure}

\begin{figure}[!h]
\centering
\includegraphics[width=3.6in]{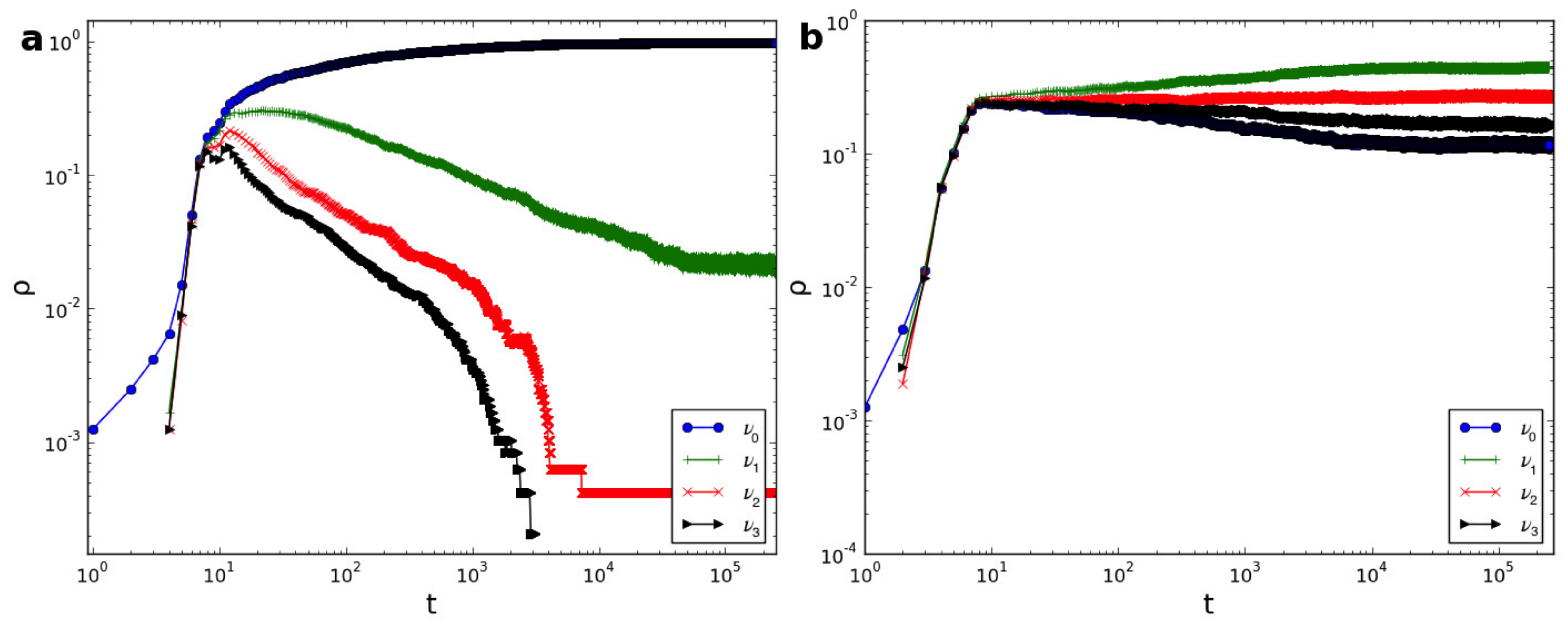}
\caption{\label{fig:density_ws_01} Density of agents who add the \linguistic, over time, in networks generated by the WS model with $\beta = 0.1$ and $N=1600$ (results are averaged over $10$ simulations). As indicated in the legend, each curve refers to a different signified, having fitness $\nu$. \textbf{a} Results for fitness set $A$. \textbf{b} Results for fitness set $C$.}
\end{figure}

\begin{figure}[!h]
\centering
\includegraphics[width=3.6in]{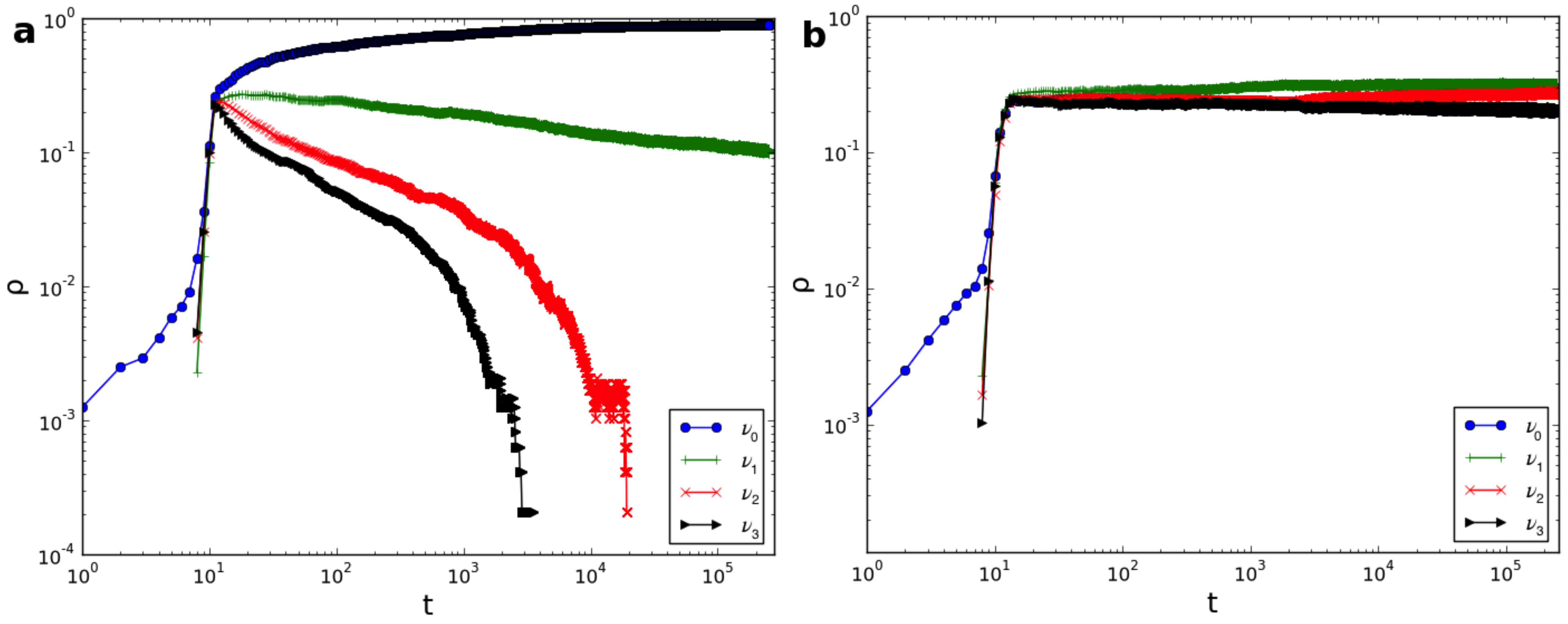}
\caption{\label{fig:density_ws_001} Density of agents who add the \linguistic, over time, in networks generated by the WS model with $\beta = 0.01$ and $N=1600$ (results are averaged over $10$ simulations). As indicated in the legend, each curve refers to a different signified, having fitness $\nu$. \textbf{a} Results for fitness set $A$. \textbf{b} Results for fitness set $C$.}
\end{figure}

\begin{figure}[!h]
\centering
\includegraphics[width=3.6in]{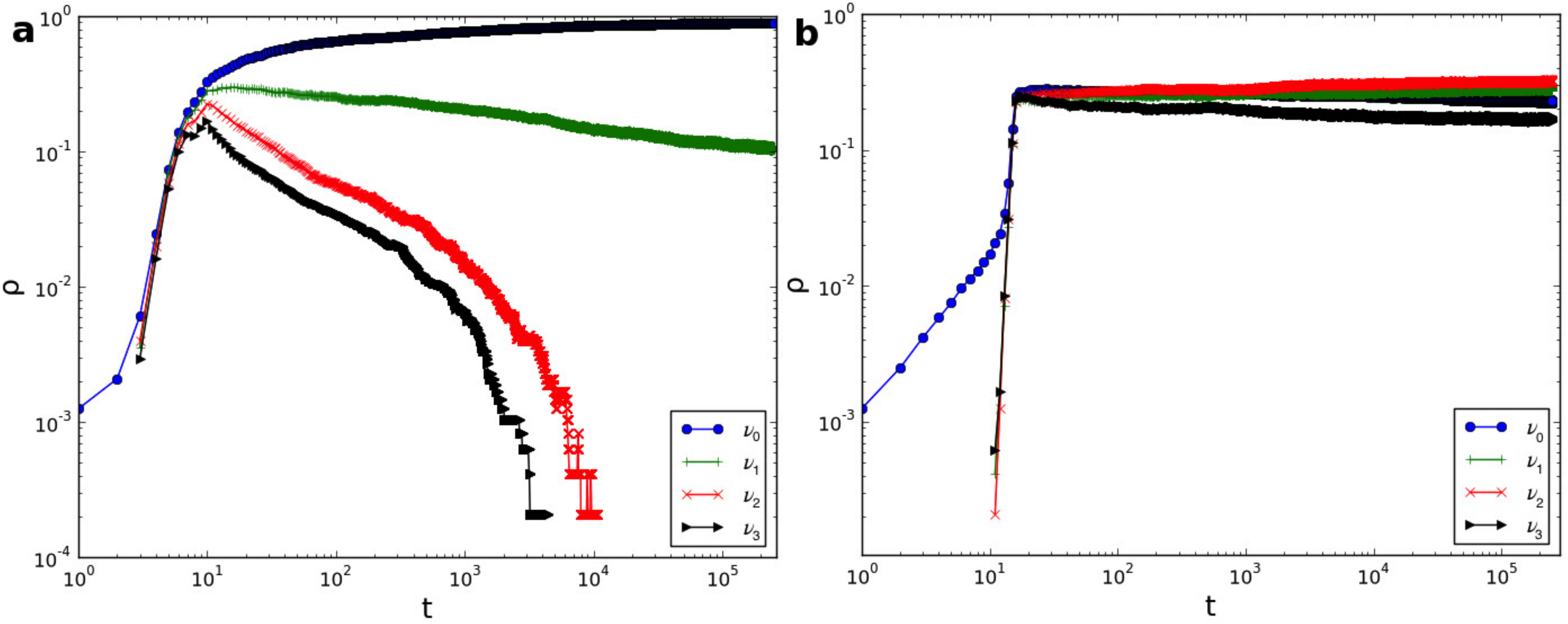}
\caption{\label{fig:density_ws_0} Density of agents who add the \linguistic, over time, in networks generated by the WS model with $\beta = 0.0$ and $N=1600$ (results are averaged over $10$ simulations). As indicated in the legend, each curve refers to a different signified, having fitness $\nu$. \textbf{a} Results for fitness set $A$. \textbf{b} Results for fitness set $C$.}
\end{figure}

Small-world networks have been generated using the Watts-Strogatz (WS hereinafter) model~\cite{watts01}. 
With the WS model it is possible to generate small-world networks by altering a regular ring lattice. In particular, the final structure of the network is obtained via a rewiring process, controlled by the parameter $\beta$. This latter is the probability of each edge being rewired, i.e., being disconnected from one of its nodes and connected to another randomly chosen node of the network. 
The parameter $\beta$ can take a value in the range $[0.0,1.0]$. In particular, the two limit cases, i.e., $\beta=0.0$ and $\beta = 1.0$, generate regular ring lattices and random networks, respectively. Instead, small-world networks are obtained for intermediate values of $\beta$. In order to compare results of WS networks, we performed simulations using $\beta = 1.0$ (see Figure~\ref{fig:density_ws_1}), $\beta = 0.1$ (see Figure~\ref{fig:density_ws_01}), $\beta=0.01$ (see Figure~\ref{fig:density_ws_001}) and $\beta = 0.0$ (see Figure~\ref{fig:density_ws_0}).
\subsubsection{Discussion}
\begin{figure}[!h]
\centering
\includegraphics[width=3.6in]{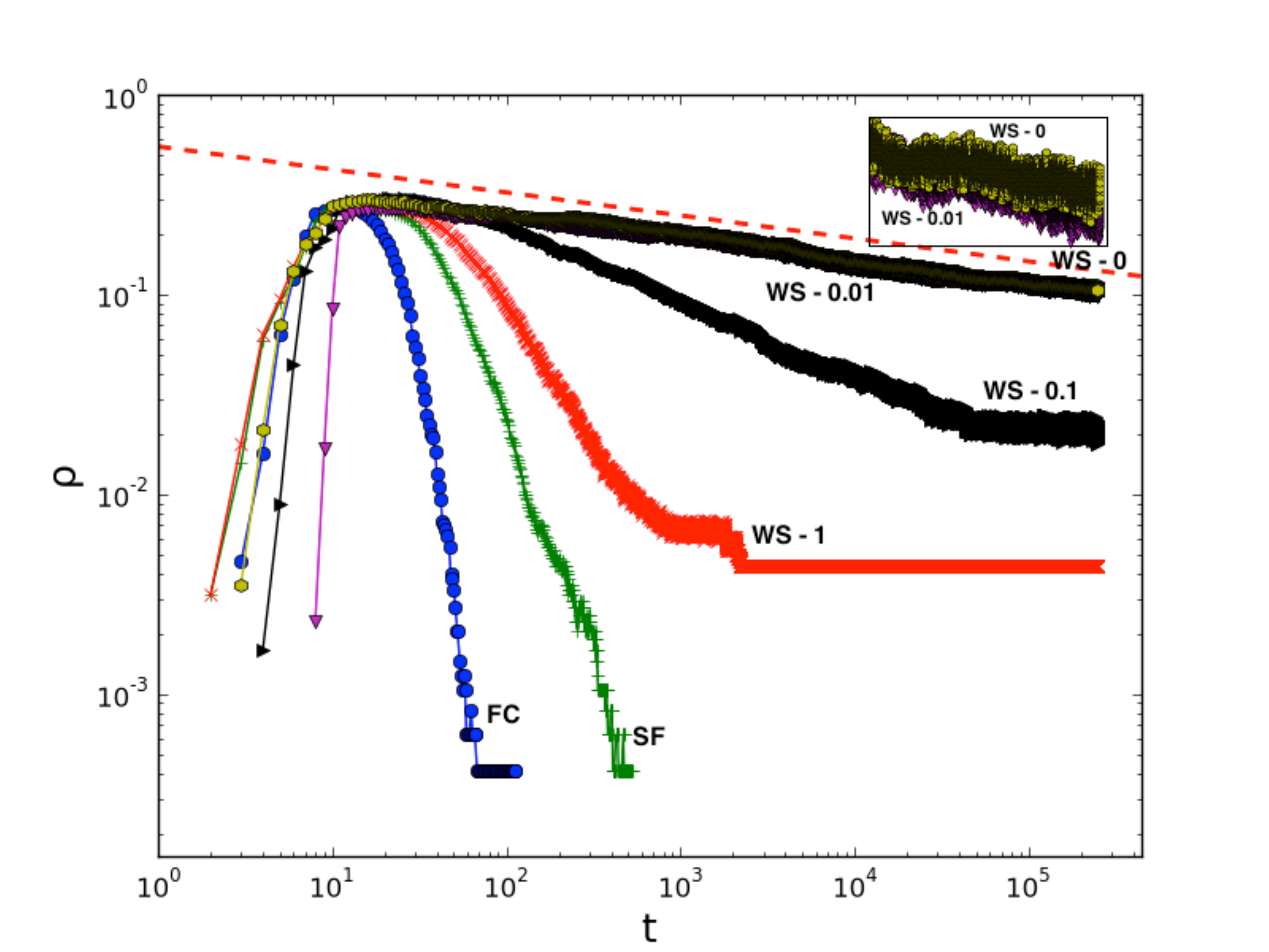}
\caption{\label{fig:density_comparative} Comparison of the density of agents adding the \linguistic~with the signified $1$ (with fitness $\nu_{1}$), using the fitness set $A$ and varying the agent network structure. The inset shows the overlap between results of WS $\beta = 0.01$ (lower) and WS $\beta = 0.0$ (upper). The dashed red line for reference: $\rho = t^{-0.12}$. Results are averaged over $10$ simulations.}
\end{figure}

It is important to examine the density of agents adding the \linguistic~to their vocabulary over time, in order to understand the process of  \linguistic~diffusion.
This analysis allows one to compare the behavior of a population confronted with different signifieds for the same sign, and to evaluate which signifieds are able to survive in relation to their fitness.
In particular, these results provide the first evidence of the importance of the fitness parameter. As expected, we found that the winning signifieds are those with the highest fitness value.
Furthermore, it is interesting to note that varying the structure of the agent network, the proposed model yields different results. For instance, simulations performed with agents organized in FC and SF networks, using the fitness set $A$ (or $B$), revealed that agents are able to converge to one common signified --see Figures~\ref{fig:density_fc} and \ref{fig:density_sf}. 
On the other hand, using WS networks and the fitness set $A$ (or $B$), more than one signified can survive --see Figures~\ref{fig:density_ws_01} and \ref{fig:density_ws_0}.
The difference in the structure of the agent networks, shown in Figure~\ref{fig:density_comparative}, can be appreciated by observing how the density of agents adding the signified $1$ (with fitness $\nu_{1}$) varies as the underlying network varies.
As denoted in Figure~\ref{fig:density_comparative} with a dashed red line, each curve $\rho$ decays following a power law $\rho(t) = t^{-\gamma}$ (note that for WS - $\beta = 1$ and WS - $\beta = 0.1$, the related curves decay until a steady state is reached).
Hence, the exponent $\gamma$ can be used as a quantitative parameter to compare the densities of agents adding the \linguistic~with the signified $1$ over time, in different agent networks -- see Table~\ref{tab:network_gamma}.
\begin{table}[!ht]
\tbl{Comparison of the exponents $\gamma$, computed for each curve related to the density of agents adding the~\linguistic~with the signified $1$, in different agent networks.}
{\begin{tabular}{ | l | l | }
\hline
Network \hphantom{00000000000000000000} &$\left \langle \gamma \right \rangle$ \hphantom{00000000000000000000} \\ \hline
WS - $\beta = 0.0$ & $0.12$ \\ 
WS - $\beta = 0.1$ & $0.38$\\
WS - $\beta = 1.0$ & $0.5$ \\
SF & $2.5$ \\
FC & $5.4$\\
\hline
\end{tabular}
\label{tab:network_gamma}}
\end{table}

\begin{figure}[!h]
\centering
\includegraphics[width=3.6in]{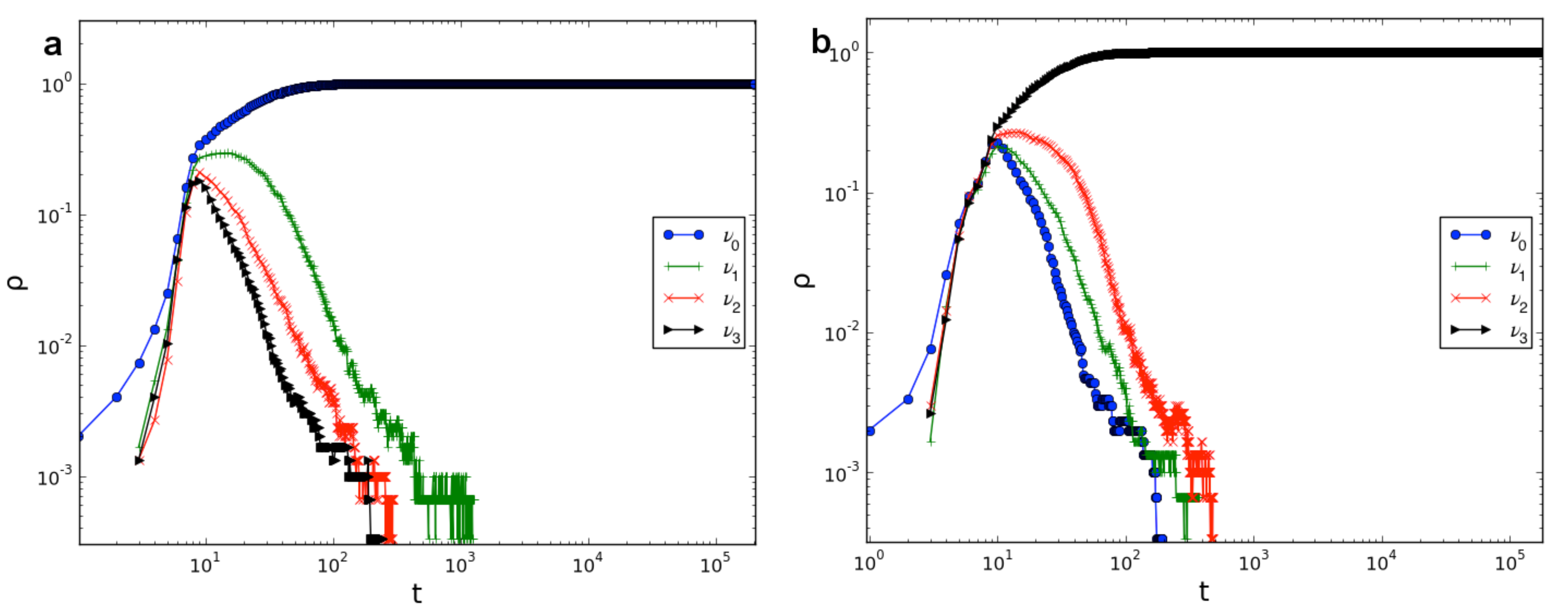}
\caption{\label{fig:density_comparison_A_B} Comparison of results obtained for FC networks with $N=1000$. On the left, simulations performed for the fitness set $A$. On the right, simulations performed for the fitness set $B$. Results are averaged over $10$ simulations.}
\end{figure}

For instance, as can be observed, the curve $\rho$ (for the signified $1$ of the \linguistic) decays much faster in SF and FC networks than in WS networks.
In particular, in FC networks the \linguistic~spreads rapidly as all agents can interact with one another. In SF networks, the diffusion  process is still very fast as a few agents have a high degree. By contrast, in the event we adopt networks generated by the WS model, the rate of diffusion diminishes substantially from random structures (i.e., WS - $\beta = 1.0$) to regular lattices (i.e., WS - $\beta = 0.0$) where the \linguistic~spreads, agent by agent, along a hubless ring.
These results demonstrate that the structure of the agent network also plays an important role in the proposed model.
All the above figures show the results obtained only for the fitness sets $A$ and $C$. Note that fitness sets $A$ and $B$ yield equivalent results. In particular, in each case, the signified with the highest and lowest fitness behave in the same way (i.e., the former survives, whereas the latter does not) --see Figures~\ref{fig:density_comparison_A_B}. 
Lastly, since the proposed model is based on two main control parameters, i.e., $\sigma$ and $\nu$, we performed a further analysis to evaluate how they influence the results obtained--see the appendices~\ref{sec:appendices}.
\subsection{Evolution of the system}
\begin{figure}[!h]
\centering
\includegraphics[width=3.6in]{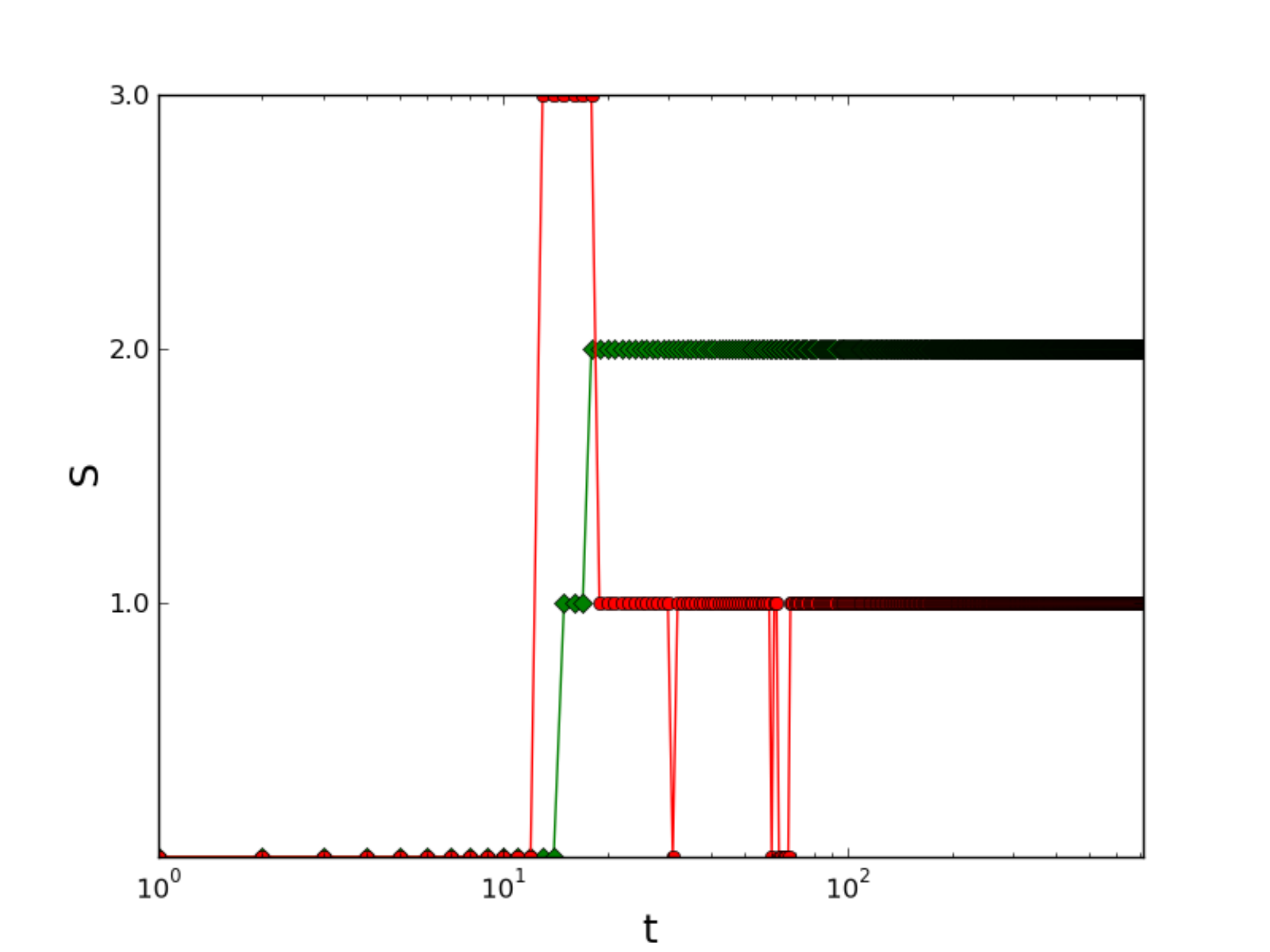}
\caption{\label{fig:transitions} Transitions between different states, undergone by two agents, during system evolution. Each color identifies an agent.}
\end{figure}

\begin{figure}[!h]
\centering
\includegraphics[width=3.5in]{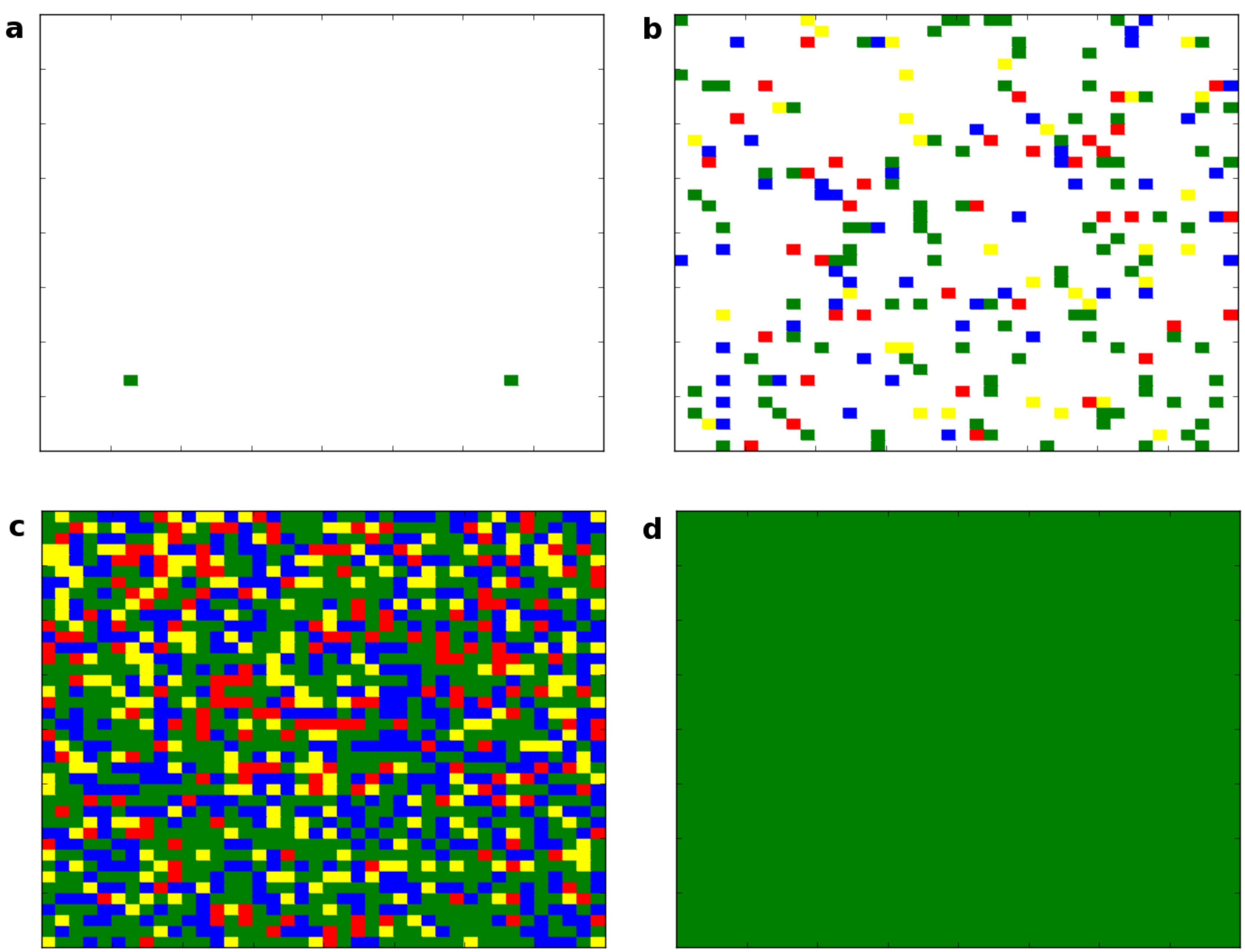}
\caption{\label{fig:evolution_fc} Evolution of the system in a FC network, for the fitness set $A$. Colors codify the state of each agent. White if the \linguistic~is unknown; green if the \linguistic~is known with its original signified (i.e., the signified introduced by its inventor); yellow and blue if the \linguistic~is known with a signified different from the original one. \textbf{a} The system at $t=1$. \textbf{b} The system at $t=5$. \textbf{c} The system at $t=1000$. \textbf{d} The system at the end of the process.}
\end{figure}

\begin{figure}[!h]
\centering
\includegraphics[width=3.5in]{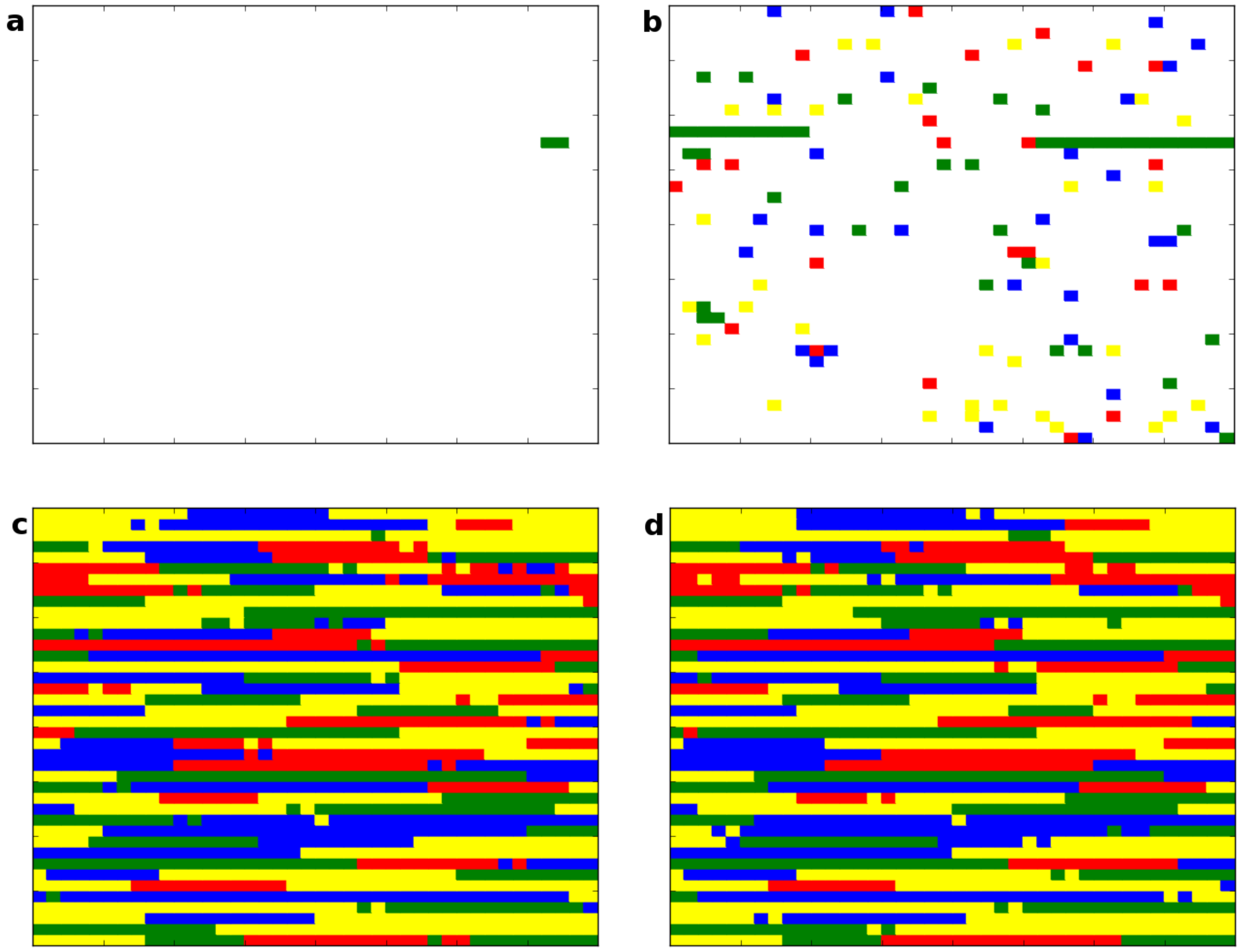}
\caption{\label{fig:evolution_ws} Evolution of the system in a WS network ($\beta=0.0$), for the fitness set $C$. Colors codify the state of each agent. White if the \linguistic~is unknown; green if the \linguistic~is known with its original signified (i.e., the signified introduced by its inventor); yellow and blue if the \linguistic~is known with a signified different from the original one. \textbf{a} The system at $t=1$. \textbf{b} The system at $t=5$. \textbf{c} The system at $t=1000$. \textbf{d} The system at the end of the process.}
\end{figure}

In the proposed model agents change their state upon interaction with their neighbors. In particular, at $t=0$, only one agent knows the \linguistic~while the others do not. This scenario can be viewed in terms of an epidemiological process~\cite{vespignani01}, where agents are in the state $S$, i.e., \textit{Susceptible}, in the event they do not know the \linguistic, in the state $I$, i.e., \textit{Infected} otherwise.
Moreover, as agents get \textit{Infected}, they can still change their state when interacting with other agents adopting a different signified for the same \linguistic. Therefore, agents change their state as follows:
\begin{equation*} \label{eq:state_system}
S \to I_x \to I_y \to I_x \to ... \to I_x
\end{equation*}
\noindent where $I_x$ and $I_y$ represent the state of an infected agent using the $x$th and $y$th signified respectively. Figure~\ref{fig:transitions} shows the transitions from one state to another, undergone by two agents, over time.
Hence, as the system evolves (i.e., the \linguistic~diffusion) agents can change their state many times.
As discussed before, agents may converge to a common signified for the \linguistic, or may not (i.e., more than one signified survives).
Figure~\ref{fig:evolution_fc} illustrates the evolution of the system when a final winning signified exists.
On the other hand, as more than one signified can survive, the final state is composed of agents in different states (obviously, all agents have been infected).
Figure~\ref{fig:evolution_ws} shows the evolution of the system, up to $2.5\cdot10^{5}$ time steps, where all signifieds survive. Therefore, as each agent can save only one signified, the network is divided into small communities.
Note that, despite the appearance of Figures~\ref{fig:evolution_fc} and \ref{fig:evolution_ws}, the structures are FC and WS with $\beta = 0.0$, respectively.
The evolution of the system shown in Figures~\ref{fig:evolution_fc} and \ref{fig:evolution_ws}, shows, in qualitative terms, the diffusion  process and the variation of the agents' state over time.
\subsection{Steady-states}
\begin{figure}[!h]
\centering
\includegraphics[width=3.7in]{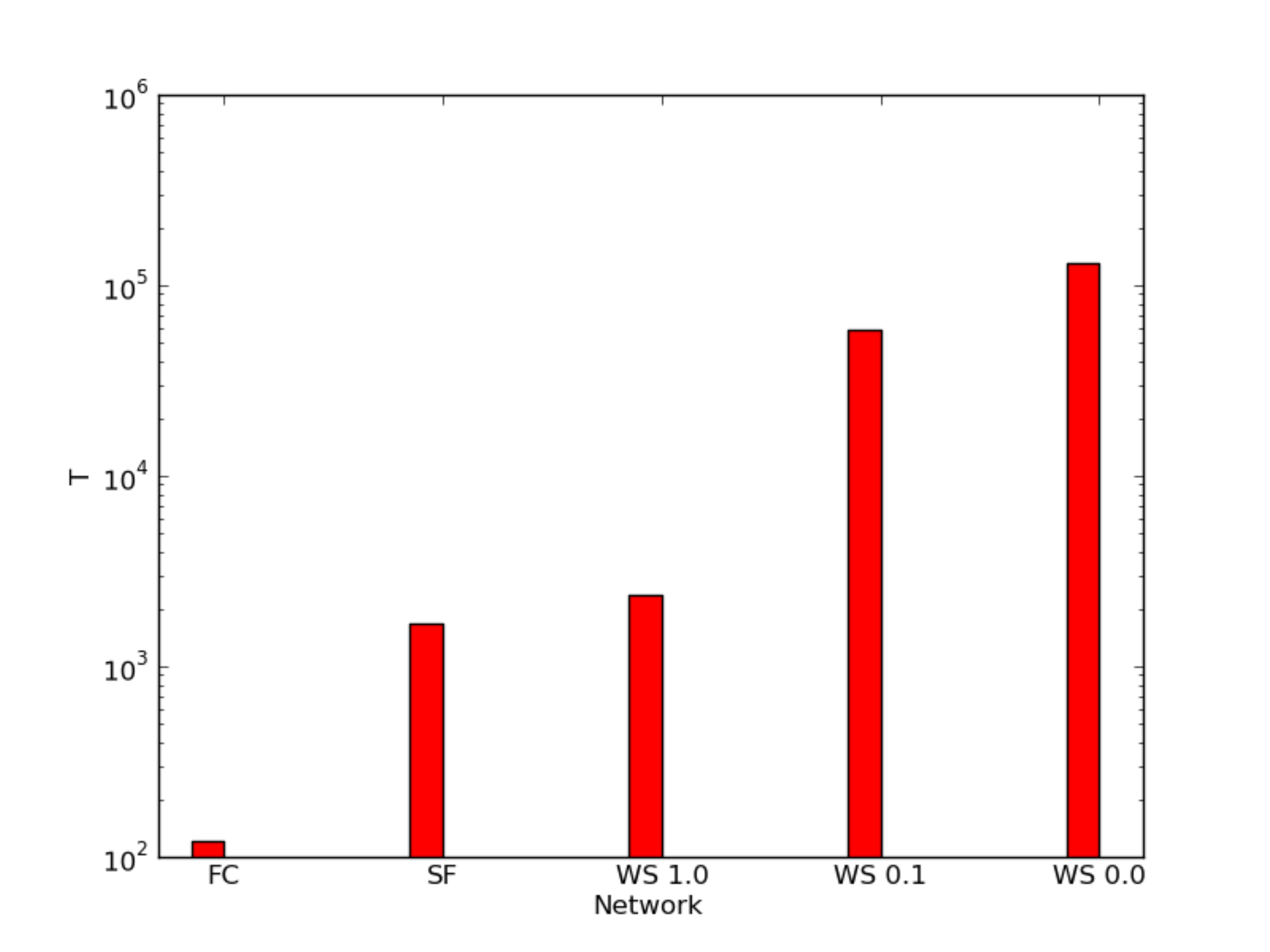}
\caption{\label{fig:time_steps} Number of time steps required to reach steady state in different agent networks with $N=1600$.}
\end{figure}

\begin{figure}[!h]
\centering
\includegraphics[width=3.7in]{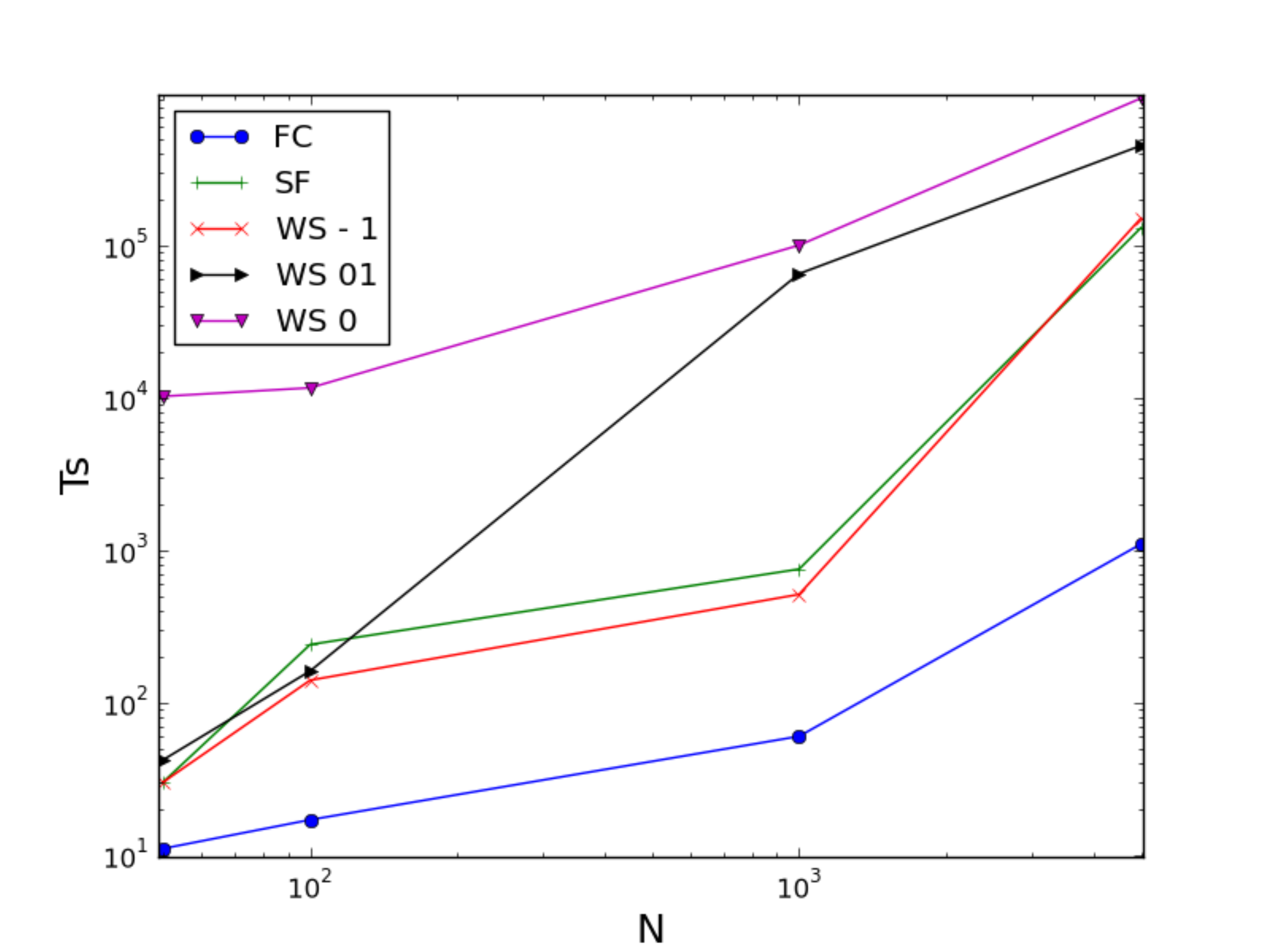}
\caption{\label{fig:time_steps_population} Number of time steps to attain steady state versus population size.}
\end{figure}

The above analysis allows one to determine whether the system attains a steady state or not, within $2.5\cdot10^{5}$ time steps. Note that this temporal limit has been introduced to ensure the algorithm terminates; otherwise in the event that more than one signified survives, the algorithm never terminates (i.e., convergence to one common signified). 
Moreover, it is interesting to compare the number of time steps required to reach the steady state (when it exists) for each agent network structure. The different structures are compared in Figure~\ref{fig:time_steps}.
As shown in Figure~\ref{fig:time_steps}, the WS networks require a larger number of time steps to reach the steady state than SF and FC networks. In particular, in FC networks agents reach the steady state in a very small number of time steps $\approx 120$ (with $N=1600$). 
Figure~\ref{fig:time_steps_population} illustrates the number of time steps needed to attain the steady state versus population size (see also \cite{wichmann01}).
As expected, the number of time steps increases with $N$. In particular, we analyzed the system varying $N$ over the range $50$ to $5000$. This analysis shows that the growth of $T_s$ (i.e., the number of time steps required to reach the steady state) is not linear. 
\subsection{Evolution of the system with unlimited  number of signifieds}
\begin{figure}[!ht]
\centering
\includegraphics[width=3.8in]{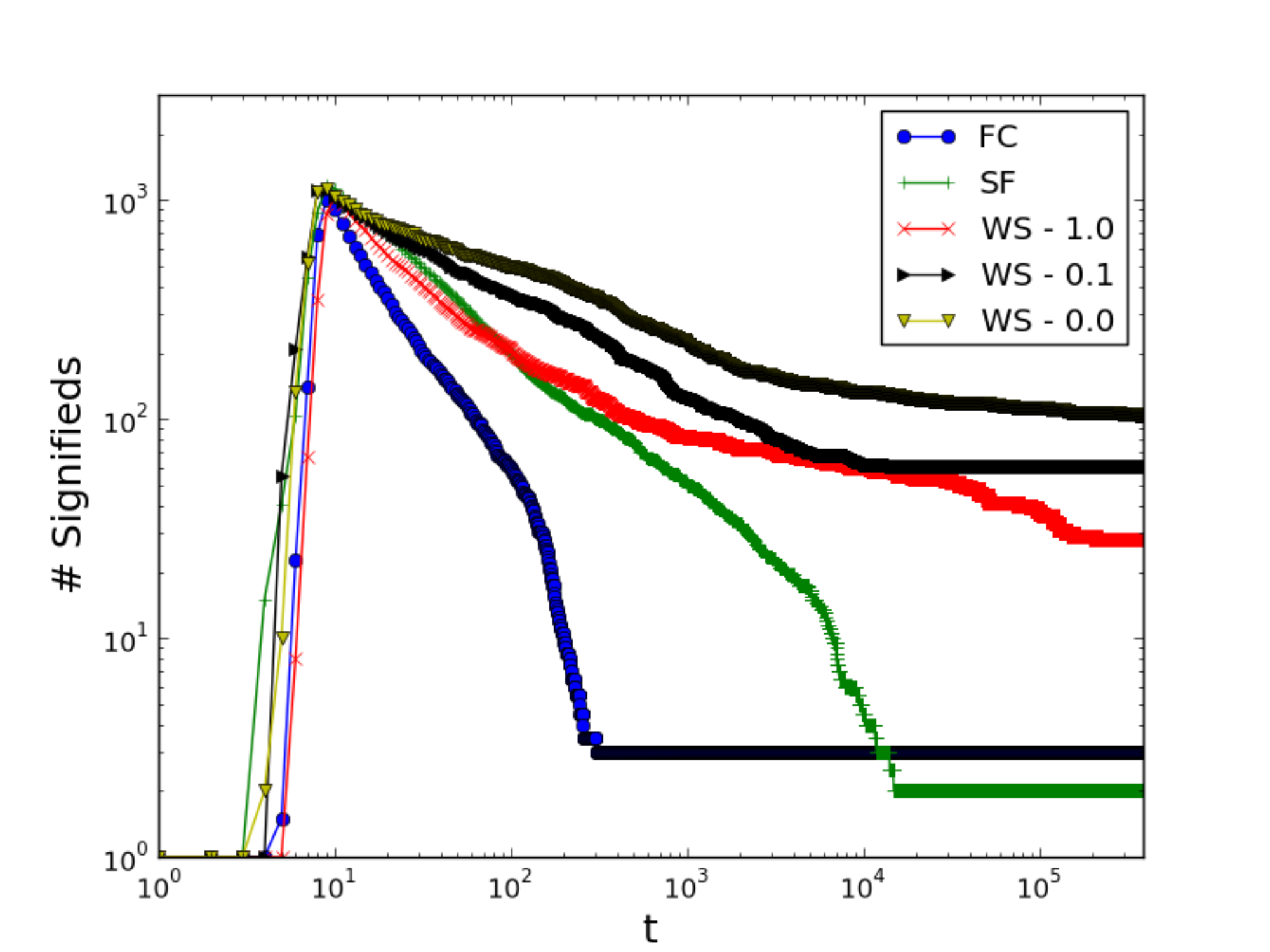}
\caption{\label{fig:evolution_relax} Number of signifieds invented by agent population over time (results are averaged over $10$ simulations). Each color identifies a network structure.}
\end{figure}

The above analysis has been conducted setting the maximum number of possible signifieds generated by agents to four.
Now, we will relax this constraint, i.e., we let the agents population generate an unlimited number of signifieds. Here too, the fitness value of each signified is randomly chosen in the range $[0,1]$. 
In order to study the unconstrained evolution of the system, we analyze the variation of the number of signifieds invented by the agents’ population over time.
Figure~\ref{fig:evolution_relax} shows the number of signifieds invented in different agent networks.
Interestingly, though the agent population defines a very large number of signifieds ($> 1000$) in a few time steps, agent interactions, driven both by the individual number of confirmations and by the fitness values, substantially reduce the final number of winning signifieds. 
In particular, within $3.5 \cdot 10^{5}$ time steps, only a few winning signifieds ($\approx 3$) survive in agent populations in FC and SF networks. 
Instead, agents in WS networks reduce the number of winning signifieds to $\approx 100$. As shown in Figure~\ref{fig:evolution_relax}, the regular ring lattice (i.e., WS - $\beta = 0$) structure yields the greatest number of winning signifieds. 
Lastly, we found that in WS networks, the number of winning signifieds decreases as $\beta$ increases. 
\section{Summary and Conclusions}\label{sec:conclusions}
The aim of this work is to analyze the dynamics of \linguistics~diffusion in communities using modern information systems. 
For the sake of simplicity, we represent this scenario as a multi-layer network comprising two layers: an agent network and a media network. The former represents a social network, whereas the latter represents the entire information system of a society (e.g., television, World Wide Web and radio). 
In order to model the behavior of \linguistics~in multi-layer networks, we use the concept of linguistic sign. We consider two main linguistic processes that a \linguistic~undergoes when introduced into a population, i.e., full reception and partial reception where new signifieds are generated. Moreover, we discuss a third linguistic process that entails generating new signifiers while saving the signified of a \linguistic~(see Appendices~\ref{sec:information_system}). 
We focus the attention on the nature of interactions between components of the two networks.
In particular, we identify two kinds of communication, i.e., ``active'' and ``passive''.  In  active communication people interact with one another immediately. Hence, in the event that a conversation between two people contains something that is unclear, the listener can immediately ask the speaker for further explanations.
Hence, in a nutshell, agents can convey both the signifier and the signified of a sign in active communication.
Instead, in passive communications, agents can only convey the signifier of a sign.
Let us assume that communication among agents is active, whereas communication between agents and media nodes is passive. Thus, agents who receive the \linguistic~via passive communication (and do not know its meaning) invent the signified associated therewith. Let us recall that, in this context, the term ``invent'' summarizes the process for identifying an appropriate signified (i.e., meaning) for the \linguistic, considering the context in which the latter is used; or for assigning a new signified although the original one is fully understood.
In so doing, interactions among agents result in all the invented signifieds, of the same \linguistic, competing to survive in the population (i.e., to be saved in the agents' vocabularies).
In particular, when two agents who know the \linguistic~interact, they use their individual number of confirmations (i.e., their experience) combined with the fitness of their signified, to decide which signified has to be saved in their vocabulary.
The results of simulations show \linguistics~to behave in a peculiar manner. In particular, agents generate numerous signifieds for the same \linguistic~and, after several interactions, only a few signifieds survive.
We found that the individual number of confirmations as well as fitness are fundamental in these dynamics.
This work brings to light the possible effects of modern information systems (see also the Appendices~\ref{sec:information_system}) on the evolution of a language and, moreover, it demonstrates the importance of the linguistic sign in modeling language dynamics.
Lastly, although the proposed model makes use of agents that are able to save only one signified (for each sign) at a time, we believe the model to be useful for representing real scenarios. In particular, we suggest that the diffusion dynamics, simulated under this constraint, can be considered as a preliminary process prior to introducing all winning signifieds into the vocabulary of the entire agent population.

\section*{Acknowledgments}
The author wishes to thank Fiorenzo Toso for his helpful comments and the Fondazione Banco di Sardegna for funding his work.
\section{Appendices}\label{sec:appendices}
\subsection{Individual number of confirmations versus Fitness}
In the proposed model, the individual number of confirmations $\sigma$ and the fitness of each signified $\nu$ play a central role. In particular, in the event two agents interact using two different signifieds of the same \linguistic, they decide which signified should be saved by Eq~(\ref{eq:signified_comparison}).
The equation has been defined to combine both parameters, i.e., $\sigma$ and $\nu$, as they are both considered important from a linguistic perspective.
Nevertheless, we are interested in evaluating the outcomes of the proposed model using these parameters singly. Moreover, this analysis also provides a rough comparison of the effects of $\sigma$ and of $\nu$.
\subsubsection*{The role of $\sigma$}
\begin{figure}[!ht]
\centering
\includegraphics[width=3.6in]{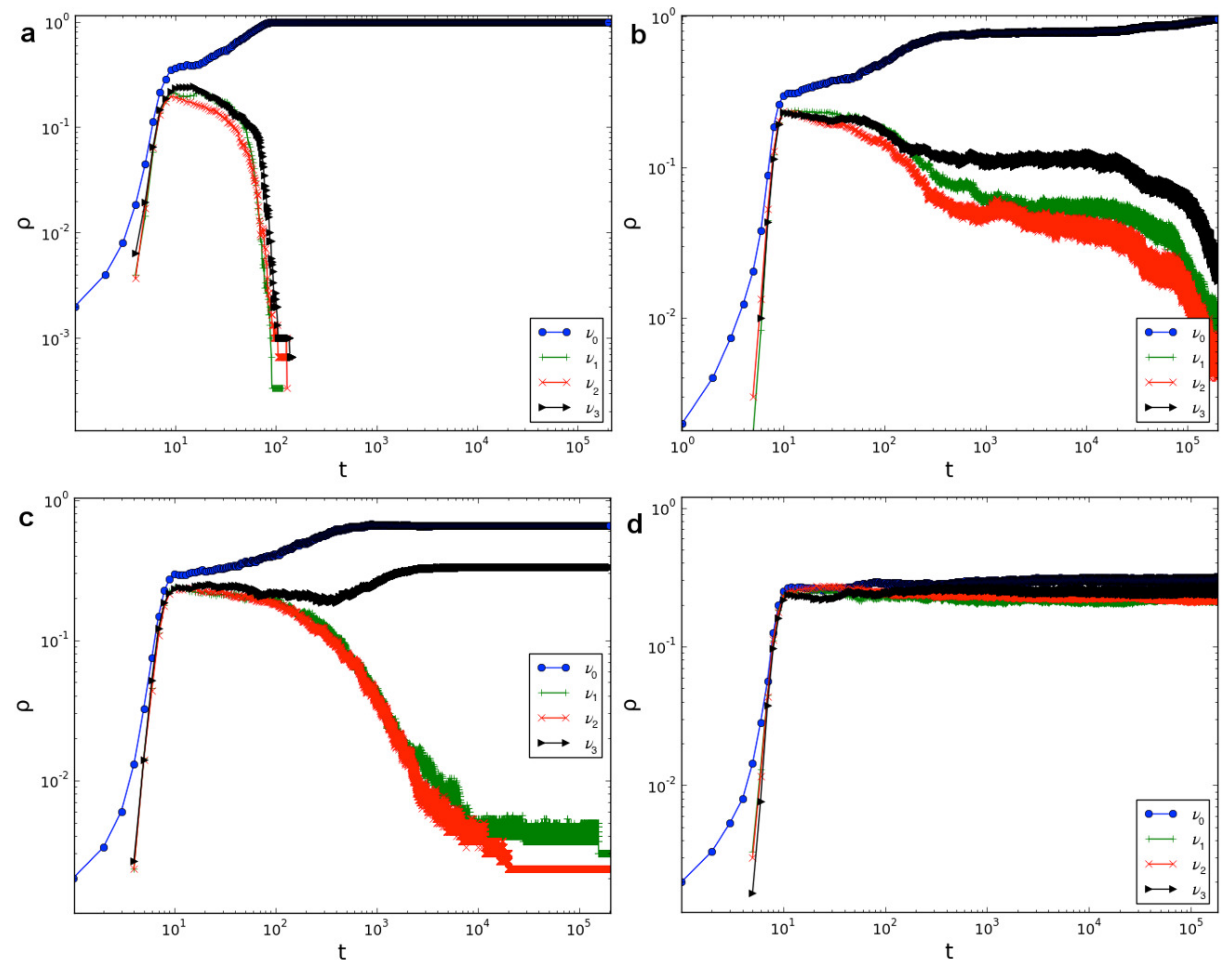}
\caption{\label{fig:density_nofitness} Density of agents adding the \linguistic over time in different agent networks, with $N=1000$ and using Eq~(\ref{eq:signified_comparison_nofitness}) (results are averaged over $10$ simulations) . As indicated in the legend, each curve refers to a different signified. \textbf{a} Results obtained for FC networks. \textbf{b} Results obtained for SF networks. \textbf{c} Results obtained for WS networks with $\beta = 1.0$. \textbf{d} Results obtained for WS networks with $\beta = 0.1$.}
\end{figure}

In order to study the role of the individual number of confirmations $\sigma$, we analyze the \linguistics~ diffusion dynamics by setting all invented signifieds to $\nu=1$. Therefore, the Eq~(\ref{eq:signified_comparison}) becomes:
\begin{equation} \label{eq:signified_comparison_nofitness}
W_x = \frac{\sigma_{x}}{\sigma_{x}+\sigma_{y}} 
\end{equation}
In so doing, interactions among agents, who know the \linguistic, are driven only by their $\sigma$ values.
As illustrated in Figure~\ref{fig:density_nofitness}, only one signified survives in FC and SF networks. In particular, the winning signified is the original one, i.e., that defined by the inventor of the \linguistic. We consider this phenomenon to be the result of a simple ``first-mover advantage'' mechanism.
On the other hand, note that when using WS networks, all signifieds survive. In WS networks generated with $\beta = 1.0$, we found two main winning signifieds and two signifieds used only by a few agents. Instead, in WS networks generated with $\beta=0.1$ (i.e., small-world) all signifieds can be considered as real winners, as the number of agents that use them is similar. 
\subsubsection*{The role of $\nu$}
\begin{figure}[!h]
\centering
\includegraphics[width=3.6in]{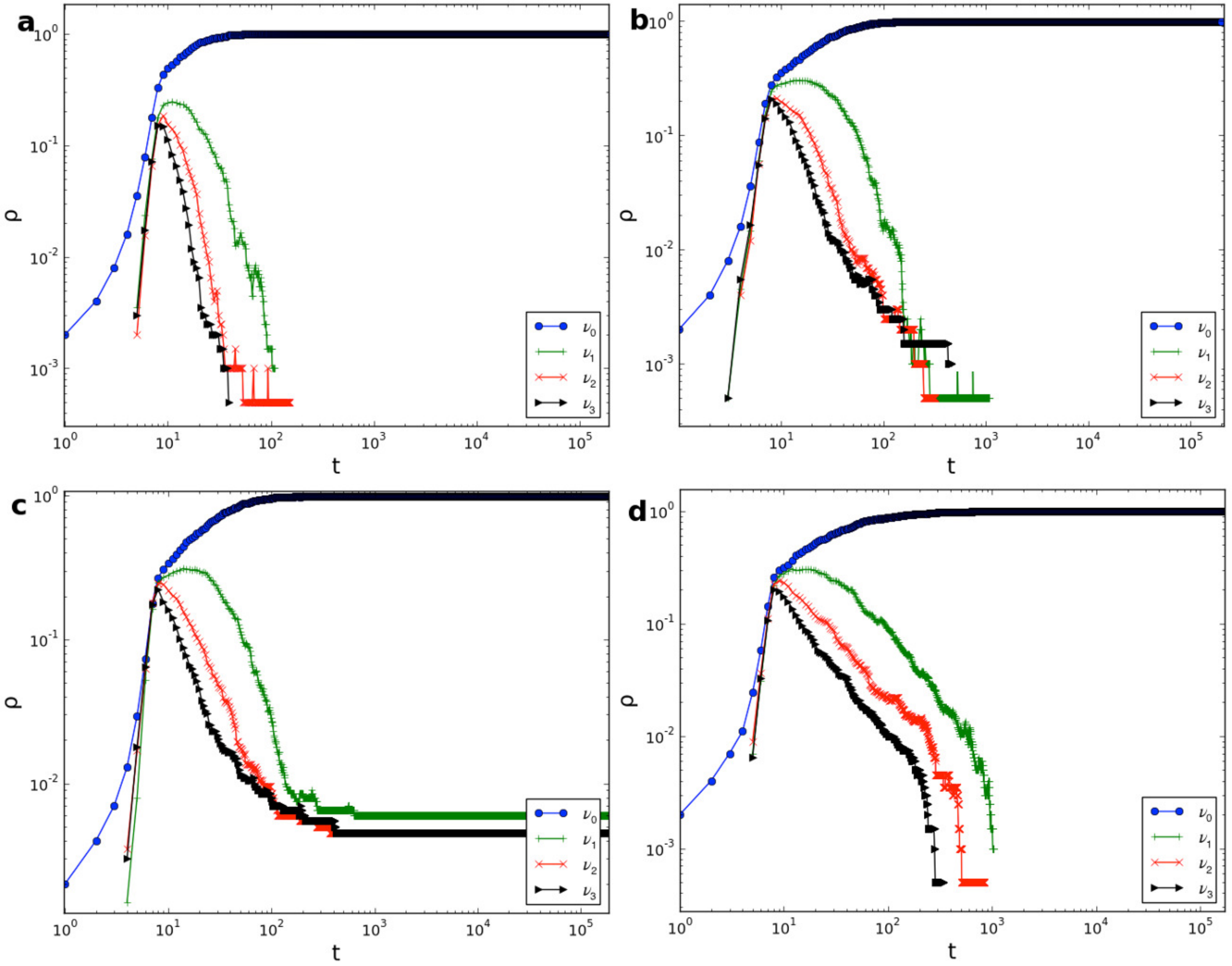}
\caption{\label{fig:density_nosigma_a} Density of agents adding the \linguistic~over time in different agent networks with $N=1000$, using Eq~(\ref{eq:signified_comparison_nosigma}) and the fitness set $A$ (results are averaged over $10$ simulations). As indicated in the legend, each curve refers to a different signified. \textbf{a} Results obtained for FC networks. \textbf{b} Results obtained for SF networks. \textbf{c} Results obtained for WS networks with $\beta = 1.0$. \textbf{d} Results obtained for WS networks with $\beta = 0.1$.}
\end{figure}

\begin{figure}[!h]
\centering
\includegraphics[width=3.6in]{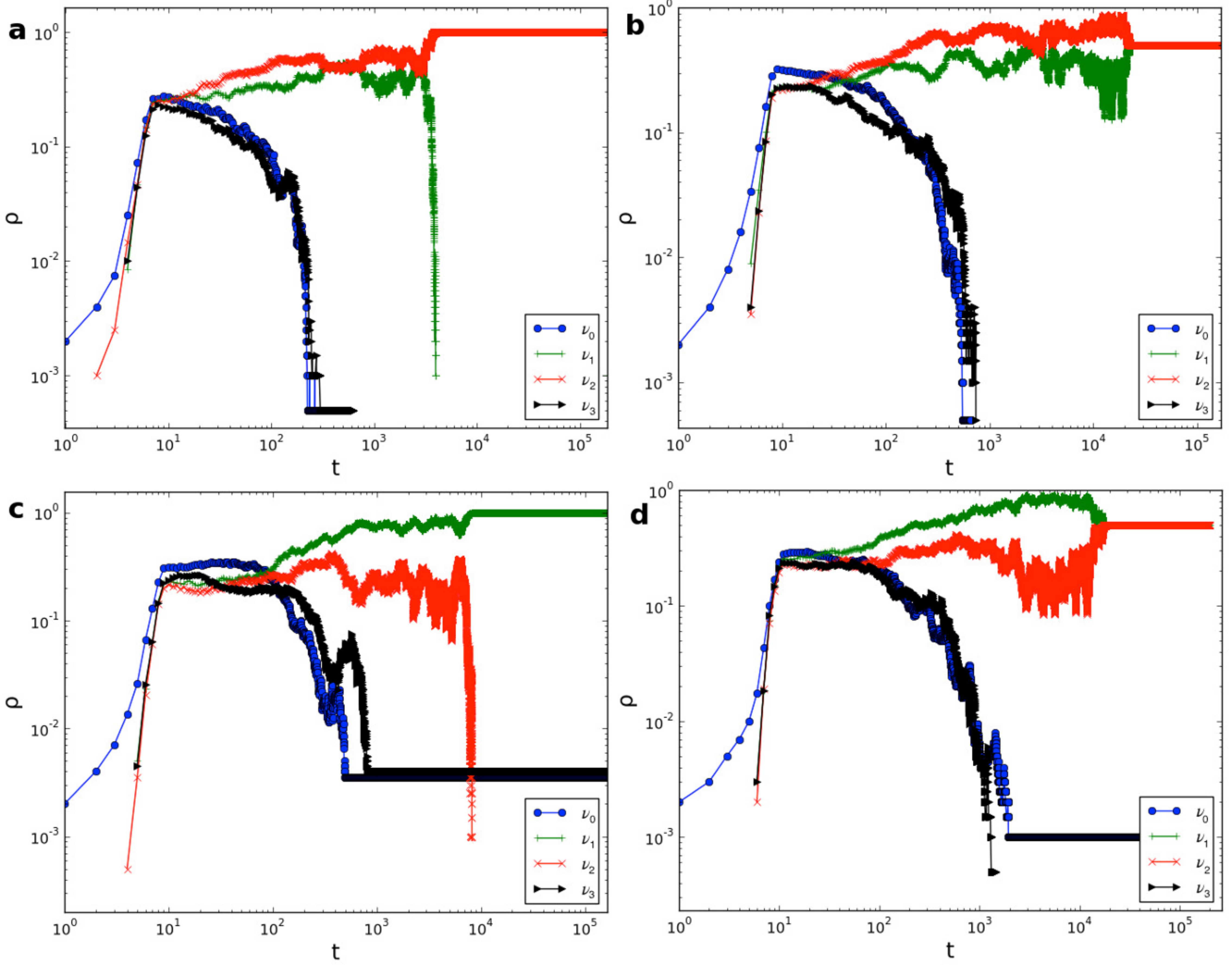}
\caption{\label{fig:density_nosigma_c} Density of agents adding the \linguistic over time in different agent networks with $N=1000$, using Eq~(\ref{eq:signified_comparison_nosigma}) and the fitness set $C$ (results are averaged over $10$ simulations). As indicated in the legend, each curve refers to a different signified. \textbf{a} Results obtained for FC networks. \textbf{b} Results obtained for SF networks. \textbf{c} Results obtained for WS networks with $\beta = 1.0$. \textbf{d} Results obtained for WS networks with $\beta = 0.1$.}
\end{figure}

To consider the effects of fitness $\nu$ alone, we analyze \linguistics~diffusion dynamics modifying Eq~(\ref{eq:signified_comparison}) as follows:
\begin{equation} \label{eq:signified_comparison_nosigma}
W_x = \frac{\nu_{k}}{\nu_{k}+\nu_{z}} 
\end{equation}
\noindent where $\nu_{k}$ and $\nu_{z}$ are the fitness values of two different signifieds, used by two agents (e.g., $x$ and $y$).
Using Eq~(\ref{eq:signified_comparison_nosigma}) the interactions among agents, who know the \linguistic, are only driven by the fitness values of their signifieds. 
Simulations have been performed using the fitness sets $A$ and $C$. As illustrated in Figure~\ref{fig:density_nosigma_a}, for the fitness set $A$ only the signified with the highest fitness survives in the population, with the exception of the case \textbf{c}, where agents are organized in a WS network with $\beta = 1.0$. Interestingly, these results are very similar to those obtained using the fitness set $A$ with Eq~(\ref{eq:signified_comparison}).
On the other hand, as shown in Figure~\ref{fig:density_nosigma_c}, the results differ from those obtained using the fitness set $C$ with Eq~(\ref{eq:signified_comparison}). In particular, in FC networks only one signified survives, in SF networks only two signifieds (those with the highest fitness) survive, and in WS networks three signifieds survives.
Note that, in the WS networks generated by $\beta=1.0$, one signified with a high fitness does not survive.
Lastly, we observe that using fitness sets such as the set $A$ (or $B$), i.e., containing different values, Eq~(\ref{eq:signified_comparison}) and Eq~(\ref{eq:signified_comparison_nosigma}) yield similar outcomes. Therefore, it seems that fitness is more important than the individual number of confirmations. 
On the other hand, using fitness sets containing similar values, such as the set $C$, the individual number of confirmations strongly affects the results, i.e., it is equally important as fitness.

\subsection{The role of the Information System}\label{sec:information_system}
\begin{figure}[!h]
\centering
\includegraphics[width=5.2in]{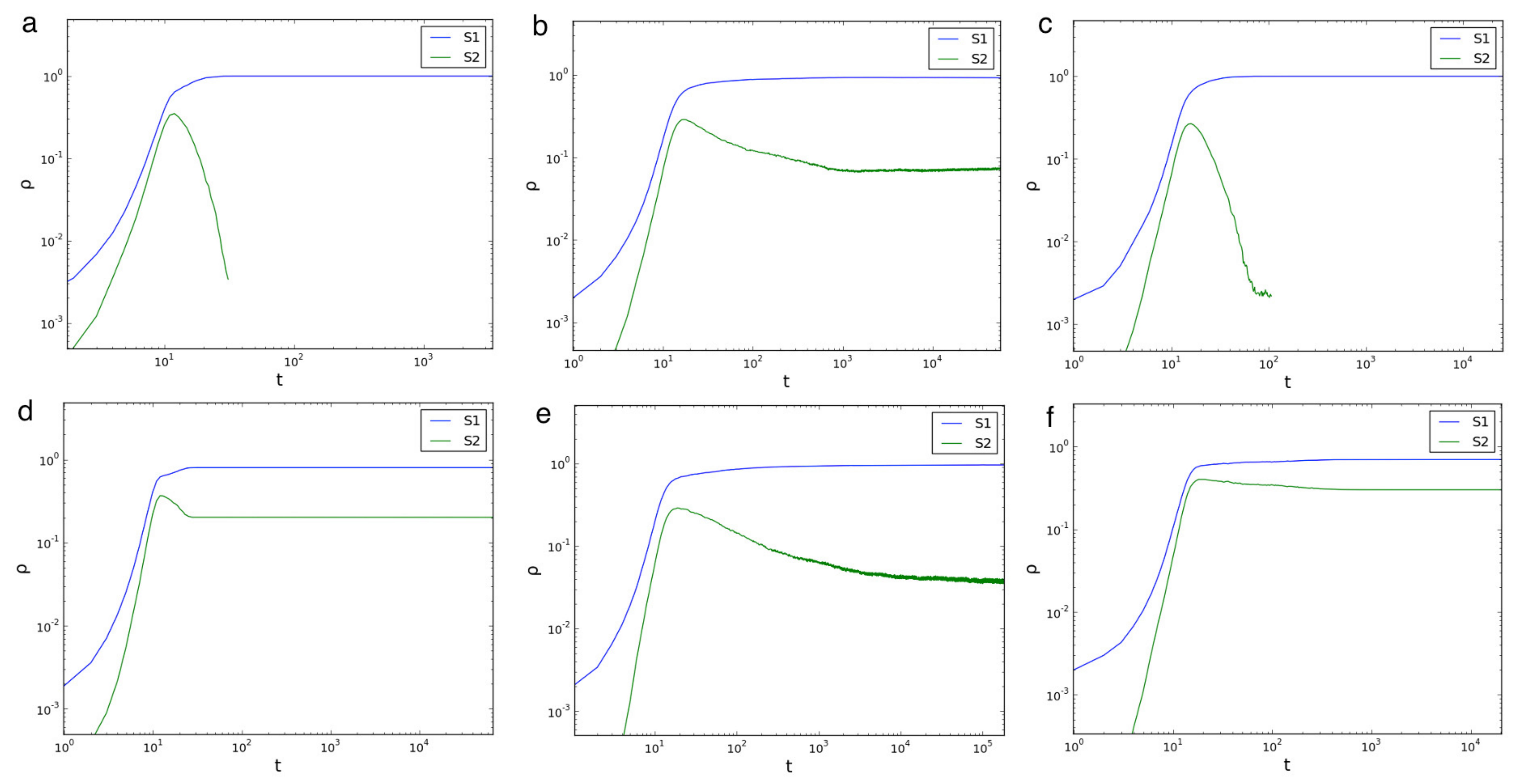}
\caption{\label{fig:information_system} Density of agents adding the \linguistic~over time, with the original signified (i.e., S1) and a second one (i.e., S2), in different agent networks with $N=1000$. The signified S1 has fitness $0.75$, whereas the signified S2 $0.5$ in the panels: \textbf{a} Results obtained for FC networks. \textbf{b} Results obtained for SF networks. \textbf{c} Results obtained for WS networks (with $\beta = 1.0$).
On the other hand, the signified S1 has fitness $0.65$, whereas the signified S2 $0.7$ in the panels: \textbf{d} Results obtained for FC networks. \textbf{e} Results obtained for SF networks. \textbf{f} Results obtained for WS networks (with $\beta = 1.0$). Results are averaged over $10$ simulations.}
\end{figure}

We briefly analyze the role of the Information System in \linguistics~diffusion dynamics. In particular, we modify the proposed model so as to consider just a single layer, i.e., the agent network. Moreover, we let agents invent only one new signified in addition to the original one.
In this case, each agent spreads the \linguistic~in the same way, i.e., sending both the signifier and the signified to its neighbors. In so doing, no misunderstandings can occur as agents only use active interactions.
Therefore, the process \textit{L1} (described in Section~\ref{sec:language}), that represents the utilization of a \linguistic~with its original signified, is still modeled; whereas the process \textit{L2}, where agents ``invent'' a new signified, is partially modeled. In particular, we introduce a small number of agents that consciously define a new signified. 
Results shown in Figure~\ref{fig:information_system} show that the probability of the second signified surviving strongly depends on fitness. In particular, in FC networks and WS networks, the second signified survives only if it has greater fitness than the original one, though the original signified always achieves wider diffusion within the population.
Interestingly, both signifieds survive in SF networks, regardless of their fitness value.
In principle, this scenario makes it possible to model the linguistic process \textit{L3}, where agents decide to save the original signified but to adopt a new signifier, i.e., in the event of a loan word, the latter is translated into the population language or into a related dialect. 
Therefore, in the light of these results we can observe that, although the information system strongly increases the rate at which the \linguistics is spread (as its nodes generate temporal links with numerous listeners/readers at the same time step), competitive signifieds also emerge and survive in its absence.

\end{document}